\documentclass[12pt]{article}

\usepackage{jeosman}
\usepackage[]{graphicx}
\usepackage{amsmath}
\usepackage{amssymb}
\usepackage{multicol}
\usepackage{epsfig}

\newcommand{\bnu}{\ensuremath{\boldsymbol{\nu}}}
\newcommand{\brho}{\ensuremath{\boldsymbol{\rho}}}

\newcommand{\bth}{\ensuremath{\boldsymbol{\theta}}}
\newcommand{\br}{\ensuremath{\mathbf{r}}}
\newcommand{\bfr}{\ensuremath{\mathbf{f}}}
\newcommand{\bu}{\ensuremath{\mathbf{u}}}

\newcommand{\comb}{\ensuremath{\text{III}}}
\newcommand{\dphi}{\ensuremath{\delta\varphi}}

\newcommand{\Dphi}{\ensuremath{\text{D}_{\varphi}}}

\newcommand{\dt}{\ensuremath{\Delta t}}

\newcommand{\fwfs}{\ensuremath{f_{\text{\tiny WFS\normalsize}}}}

\newcommand{\lpg}{\ensuremath{g_{\text{\tiny loop}}}}
\newcommand{\gdm}{\ensuremath{\gamma_{\text{\tiny DM}}}}
\newcommand{\Gdm}{\ensuremath{\Gamma_{\text{\tiny DM}}}}
\newcommand{\lwfs}{\ensuremath{\Lambda_{\text{\tiny WFS}}}}
\newcommand{\maskl}{\ensuremath{\mu_{\text{\tiny LF}}}}
\newcommand{\maskh}{\ensuremath{\mu_{\text{\tiny HF}}}}
\newcommand{\mphi}{\ensuremath{\overline{\varphi}}}

\newcommand{\sinc}{\ensuremath{\text{sinc}}}

\newcommand{\tlag}{\ensuremath{t_{\text{lag}}}}

\newcommand{\otfsys}{\ensuremath{\mathrm{OTF}_{\mathrm{SYS}}}}
\newcommand{\otfado}{\ensuremath{\mathrm{OTF}_{\mathrm{AO}}}}

\newcommand{\otftel}{\ensuremath{\mathrm{OTF}_{\mathrm{TSC}}}}
\newcommand{\eq}[1]{Eq. (\ref{#1})}

\begin{document}

\title{Synthetic Modeling of Astronomical Closed Loop Adaptive Optics}

\author{Laurent Jolissaint\footnote{now at laurent.jolissaint@aquilaoptics.com}}

\address{Leiden Observatory, Niels Bohrweg 2, NL-2333 CA Leiden, The Netherlands}

We present an analytical model of a single natural guide star astronomical adaptive optics system, in closed loop mode. The model is used to simulate the long exposure system point spread function, using the spatial frequency (or Fourier) approach, and complement an initial open loop model. Applications range from system design, science case analysis and AO data reduction. All the classical phase errors have been included: deformable mirror fitting error, wavefront sensor spatial aliasing, wavefront sensor noise, and the correlated anisoplanatic and servo-lag error. The model includes the deformable mirror spatial transfer function, and the actuator array geometry can be different from the wavefront sensor lenslet array geometry. We also include the dispersion between the sensing and the correction wavelengths. Illustrative examples are given at the end of the paper.

\keywordline astronomical instrumentation, adaptive optics, Fourier optics modeling, spatial frequency modeling

\section{INTRODUCTION}

\begin{figure}[p]
   \centering
   \epsfig{file=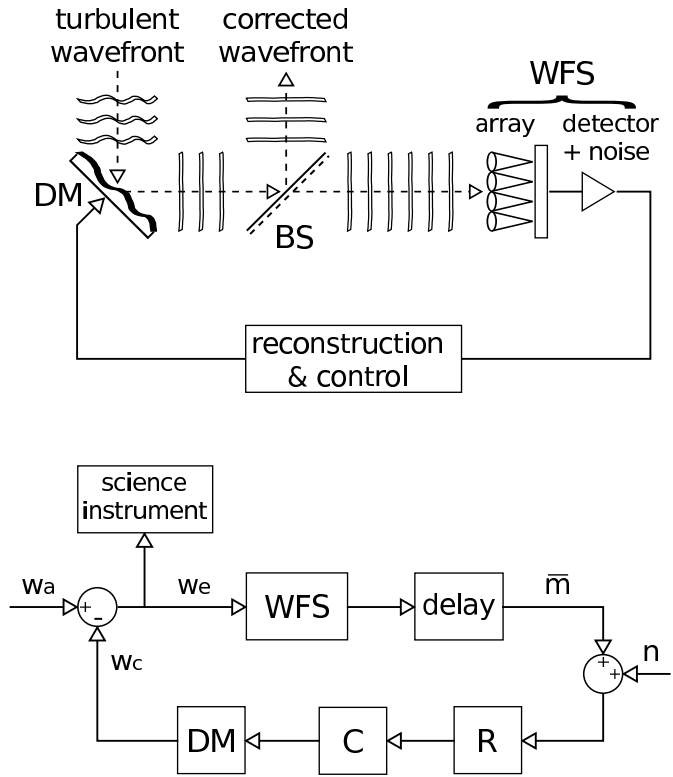,width=8.7cm}
   \caption{Top: basic elements of an AO closed loop system. The incoming turbulent wavefront is transmitted by the telescope optics to the entrance of the AO system. The deformable mirror surface shape (DM) is set by the control computer to compensate for the wavefront error. A fraction of the corrected light beam is transmitted by the beam-splitter (BS) to the wavefront sensor (WFS) where the residual wavefront is measured. The computer reconstruct the residual wavefront from the WFS measurement and computes/updates the DM surface to keep the wavefront residual as small as possible. The other part of the corrected beam is sent toward the science instrument. Bottom: equivalent block diagram of the AO closed loop system, indicating the main elements of the loop. R indicates the wavefront reconstruction operation, C is the control algorithm (an integrator in this paper).}
   \label{fig:01}
\end{figure}

Astronomical adaptive optics (AO) are complex opto-electro-mechanical systems, designed to correct random aberrations generated by optical turbulence in earth's atmosphere, and improve the performance of astronomical telescopes - see Roddier et al. \cite{roddier:99} and figure \ref{fig:01} for an illustration. Modeling the performance of such a system - for science cases analysis, AO data reduction and system design - requires sophisticated simulations tools. The most intuitive approach to construct an AO simulation tool is to break-down the system in its fundamental components, build a physical model for each of these components, and link all of them following a block-diagram architecture. The turbulent phase is then propagated through the model, mimicking a real system, up to the system's output, the focal plane, where the system's point spread function (PSF) is measured. The capacity of such end-to-end models to produce accurate performance predictions is in principle limited only by our capability to accurately code the behavior of each sub components and the system's input disturbance (optical turbulence, noise, other) - see for instance Carbillet et al. \cite{carbillet:05} and Le Louarn et al. \cite{lelouarn:06}.

End-to-end models have a limitation, though. Indeed, as the input of the system (optical turbulence) is of stochastic nature, the model needs to be run on a large number of instantaneous turbulent optical waves, typically more than a thousand to converge to a pseudo-long exposure PSF from which performance metrics can be estimated. Such a procedure takes a lot of time: depending on order of the AO system, which defines the size of the command matrix, and the level of sophistication of the end-to-end model, getting a PSF without too many residual speckles can take several hours or even days. As a consequence, performance analysis using end-to-end tools is generally limited to a few well selected representative cases, and extensive studies of the system parameter space is rarely undertaken. Also, because all the error sources are naturally merged in an end-to-end model, it is not easy to disentangle the impact of the individual sources of error on the overall performance, unless one has a deep understanding of how the end-to-end model was built. End-to-end models are therefore not a good choice for AO engineers for the broad analysis of a system performance, nor for astronomers interested in exploring the capability of a given AO system (existent or planned) for their science programme.

To suppress the limitation of end-to-end models, Rigaut et al.\cite{rigaut:98} proposed a totally different method, that we refer to here as the synthetic approach. Instead of letting the optical wave propagate through the system components and observe what comes at the output, we built a model for the system's output itself. The synthetic approach is based on an understanding of the system's behavior, and its accuracy is only limited by this understanding. \textit{A priori} knowledge is critical for the synthetic approach, while this is not needed with the end-to-end approach. We can also said that the synthetic approach models the \textbf{behavior} of the system, while the end-to-end approach models the \textbf{structure} of the system.

The starting/central point in the synthetic approach is the construction of an analytical model for the long exposure (or average) AO-corrected phase spatial power spectrum density (s-PSD). AO correction is actually seen as a spatial filter applied on the turbulent phase s-PSD. This approach is therefore also called the Fourier method or spatial frequency method in the AO literature. From this residual phase s-PSD it is shown in this paper how the long exposure PSF can be computed, in a few steps. Getting the long exposure PSF is therefore very fast and exploring in detail the AO system parameter space becomes possible. Also, wavefront error budgets are easily build, because a s-PSD is attributed to each error source, from which we can get the wavefront error variance, by numerical integration in the spatial frequency domain. Finally, by nature, there are no residual speckles in a synthetic PSF: performance metrics (Strehl ratio, PSF width, integrated energy) are therefore not "noisy", and science case analysis are greatly facilitated.

This being said, the synthetic approach has its limits: non linear effects cannot be modeled, neither transient temporal behaviors. Also, the fundamental assumption is that the corrected phase is stationary within the pupil (which is not true near the edge of the pupil and for low order aberrations) and this can produce pessimistic performance predictions.  For these reasons, end-to-end and synthetic models are to be seen as complementary rather than competitive methods: broad exploration of the system's performance is the domain of synthetic models, while detailed analysis of specific aspects of the system definitely requires end-to-end models. Actually, inclusion of our synthetic model within an end-to-end Monte-Carlo code has already been tried successfully by Carbillet et al.\cite{carbillet:10}.

In an earlier work \cite{jolissaint:06}, we complemented Rigaut et al.'s work, explaining the foundations of the method in great details, and including PSF modeling in two dimensions. This initial model was open loop, where the wavefront sensor (WFS) measures the wavefront aberration \textit{before} the deformable mirror (DM) correction, while the vast majority of current systems operate in a closed loop mode, i.e. the WFS measures the residual wavefront error \textit{after} compensation by the DM - see figure \ref{fig:01}.

Closed loop systems control the wavefront error and are therefore relatively insensitive to external disturbances (like noise) and internal variations of system parameters. Open loop systems on the contrary are very sensitive to modeling errors: it is critical that the system's components behave the way they are supposed too, as the quality of the correction is not controlled. On the other hand, feedback of the wavefront error in a closed loop system can generate diverging instabilities, if the error is overcompensated, or if time delay in the loop is too large. Open loop systems do not have this stability issue. See Ogata \cite{ogata:97} for an introduction on control systems.

In bright guide star conditions, the AO system can be run at a high loop rate: the servo lag error (due to the time lag between the measurement and the actual correction) is low, and the noise level is negligible. In this case there are no differences between open and closed loop performance. For a dim guide star, the WFS exposure time is increased to gather more photons and keep the signal/noise ratio of the wavefront error measurement at an acceptable level: servo-lag error increases, and because of differences in open and closed loop transfer functions, the system performance significantly changes between the two modes. Because of this different behavior, open loop models cannot be used to predict close loop performance in high noise regime. As predicting the limiting magnitude of a system is of great importance, in particular for science cases studies, we have developed further our initial model and included closed loop modeling.

We also took this opportunity to introduce a DM spatial frequency transfer function, allowing the analysis of different influence function structures and actuators grid architecture, and the dispersion error, i.e the error induced by the air's refractive index dispersion, which makes that the wavefront measured at a given wavelength is slightly different from the wavefront corrected at any other wavelength, generating a non negligible error for tight error budget AO systems. Also, a few conceptual errors that appeared in our initial open loop paper are corrected.

Our closed loop model is developed for a single natural guide star (NGS) Shack-Hartmann WFS based system (SH-WFS), and a least square error (LSE) wavefront reconstruction algorithm. This case covers the vast majority of current NGS-based AO systems design, and in any case our model can be easily adapted to other schemes. Examples of usage of the synthetic method are given at the end of this paper, and we show how this method can be used to optimally dimension a system.

We finally note that the synthetic approach has been used and developed by other authors as well, increasing the diversity of views and understandings of the limits and strengths of this method. We must mention first the work of Ellerbroek \cite{ellerbroek:05} who basically developed the same method using a more general albeit potentially less detailed approach; the work of Rigaut\cite{rigaut:01}, Tokovinin \cite{tokovinin:04} and Jolissaint et al. \cite{jolissaint:05} for a ground-layer AO mode (but not in closed loop and without wavefront sensor noise model); and more recently Neichel et al. \cite{neichel:08} who introduced multiple guide star tomographic reconstruction, a very useful extension of the method. What our model brings to these recent developments is essentially the closed loop mode, and some useful sophistications like the DM transfer function. To finish, it is fair to mention that R. Conan and Ch. Verinaud (private communication) both independently developed a synthetic closed loop model, yet unpublished, using another but equivalent approach than the one presented here, namely the equivalence between the spatial and temporal frequency through the Taylor hypothesis (see text).

\section{FOUNDATIONS OF THE SPATIAL FREQUENCY METHOD: A SUMMARY}\label{sec:section2}

A detailed description of the foundations of the method is given in Jolissaint et al.\cite{jolissaint:06}. A summary is given here for convenience.

The method is based on the relationship between the phase spatial frequency power spectrum and the phase structure function (SF) in one side, and the SF and the average (long exposure) AO system optical transfer function (OTF) on the other\footnote{remember that OTF and PSF are Fourier transforms of each other}. Let us review the different steps from the phase s-PSD to the long exposure PSF.

Analytical expressions for the s-PSD will be developed later. The starting point is the OTF of the whole system made of (1) the column of air above the telescope, seen as an optically transparent medium with a turbulent field of refractive index, (2) the telescope optics, possibly with static aberrations, (3) the AO system optics, (4) the science instrument optics, that can be merged with the telescope optics. Splitting the phase aberration into a static, constant part \mphi\ and a turbulent, zero mean part \dphi,
\begin{equation}
\varphi(\br,t)=\mphi(\br)+\dphi(\br,t)
\label{eq:01}
\end{equation}
where \br\ is the position vector in the pupil plane and $t$ is the time, and remembering that the OTF is also given by the autocorrelation of the phasor $\exp{(i\varphi)}$ in the pupil plane \cite{goodman:96}, we get, for the long exposure OTF (averaged over an infinite number of realization of the random AO corrected turbulent phase)
\begin{multline}
\otfsys(\bnu)=\frac{1}{S_{p}}\iint\limits_{\mathbb{R}^2}
\big\langle\exp\big\{i[\dphi(\br,t)-\dphi(\br+\brho,t)]\big\}\big\rangle_{t}\times\\
\exp\big\{i[\mphi(\br)-\mphi(\br+\brho)]\big\}\text{P}(\br)\text{P}(\br+\brho)\,\text{d}^{2}r
\label{eq:02}
\end{multline}
where \bnu\ is the angular frequency vector in the focal plane, associated to the spatial shift $\brho=\lambda\bnu$ in the pupil plane ($\lambda$ is the optical wavelength), $S_{p}$ is the pupil area and $\text{P}(\br)$ is the pupil transmission (1 inside the pupil, 0 outside), and $\langle\cdot\rangle$ indicates a time or ensemble average. Assuming a Gaussian statistics for the phase aberration, which is a very good assumption for the uncorrected as well as for the AO corrected phase, it is shown in Roddier \cite{roddier:81} that the average can be moved into the exponential argument, and we get
\begin{multline}
\otfsys(\bnu)=\frac{1}{S_{p}}\iint\limits_{\mathbb{R}^2}
\exp\big[-1/2\,\Dphi(\br,\brho)\big]\times\\
\exp\big\{i[\mphi(\br)-\mphi(\br+\brho)]\big\}\text{P}(\br)\text{P}(\br+\brho)\,\text{d}^{2}r
\label{eq:03}
\end{multline}
where
\begin{equation}
\Dphi(\br,\brho)\equiv\langle[\dphi(\br,t)-\dphi(\br+\brho,t)]^2\rangle_{t}
\label{eq:04}
\end{equation}
defines the phase structure function, a measure of the variance of the phase difference between two points separated by a vector \brho\ in the pupil. We see that the structure function depends not only on the separation vector \brho, but also on the location \br\ where the phase difference is measured. Therefore, if we want to compute the long exposure OTF, we need a model in (\br,\brho) of the structure function, which is not necessarily difficult to obtain, analytically, but what is more annoying is that we need then to perform a numerical integration of \eq{eq:03} over \br\ for each angular frequency $\bnu=\brho/\lambda$. This can be a very time consuming effort, and goes against the very objective of the synthetic approach.

Now, it is demonstrated in \cite{roddier:81} that the optical turbulence phase - {\it before} AO correction - is stationary over the pupil, i.e. its statistical properties do not depend on the location \br\ inside the pupil. The corrected phase, on the other hand, is not stationary, and its residual variance increases from the center to the edge of the pupil. This being said, this non stationarity affects mostly the first orders - piston, tip-tilt, defocus ... - and the corrected phase can be considered to be stationary for a moderate to high order AO system (i.e. moderate to high Strehl). If the phase is stationary, which we will assume from now on, the phase structure function can be written $\Dphi(\br,\brho)=\Dphi(\brho)$ and the structure function exponential can be extracted from the integral in \eq{eq:03}, so we get
\begin{multline}
\otfsys(\bnu)\approx\exp{\big[-\Dphi(\brho)/2\big]}\,
\frac{1}{S_{p}}\iint\limits_{\mathbb{R}^2}
\exp\big\{i\big[\mphi(\br)-\mphi(\br+\lambda\bfr)\big]\big\}
\text{P}(\br)\text{P}(\br+\lambda\bfr)\,\text{d}^{2}r\\
=\otfado(\bnu)\,\otftel(\bnu)
\label{eq:05}
\end{multline}
The system's OTF can therefore be seen as the telescope OTF filtered by an AO OTF filter, $\exp[-\Dphi(\brho)/2]$. Separating the OTF this way is equivalent to splitting the overall optical system into two independent systems: the optical turbulence plus AO system on one side, and the telescope plus instrument optics on the other. The first system is therefore not related to the pupil optics in any way, and its description does not include the pupil boundaries anymore. For this very reason, synthetic modeling is sometime referred to as \textit{infinite aperture} modeling.

We discuss now the relationship between the stationary SF and the phase s-PSD. Thanks to the stationary assumption, the AO system can be considered as an optical system applying a correction on a turbulent phase, regardless of any beam boundaries, all over an hypothetical plane perpendicular to the telescope optical axis. Everything looks as if the phase was pre-corrected by the AO system \textit{before} being intercepted by the telescope beam. Now, it is shown in \cite{tatarski:61} that the stationary structure function \Dphi\ is related to the spatial correlation of the phase, $\text{B}_{\varphi}$, which is itself equal to the Fourier transform of the phase s-PSD (written $\Xi_{\varphi}$ here), and we get
\begin{equation}
\Dphi(\brho)=2\big[\text{B}(0)-\text{B}(\brho)\big]=
2\iint\limits_{\mathbb{R}^2}\big[1-\cos{(2\pi\bfr\brho)}\big]\,\Xi_{\varphi}(\bfr)\,\text{d}^2\!f
\label{eq:06}
\end{equation}
The integral of the phase s-PSD gives the phase variance, and the cosine term is there instead of the usual complex exponential we have in a Fourier transform, because as the phase s-PSD is even, the sine component of the FT would be naturally null. With this last equation, we have completed the link between the phase s-PSD and the long exposure PSF.

To summarize, the procedure to get the long exposure PSF from the phase s-PSD is the following:
\begin{enumerate}
\item the phase s-PSD is computed from the analytical expressions given in the next sections, for each of the wavefront error components,
\item the phase SF is computed from the phase s-PSD with \eq{eq:06}, using a numerical Fourier transform algorithm,
\item the AO-OTF filter is given by the exponential of the SF, $\exp{(-\text{SF}/2)}$,
\item the telescope OTF is computed analytically or numerically and is filtered by the AO-OTF,
\item the long exposure PSF is obtained by applying a numerical Fourier transform on the final OTF.
\end{enumerate}
The whole procedure therefore consists in the evaluation of a few analytical expressions and the computation of two numerical Fourier transforms.
 
\section{SPATIAL FREQUENCY POWER SPECTRUM OF THE AO CORRECTED PHASE}

\subsection{The fundamental equation of adaptive optics}

The starting point for the development of the phase s-PSD is the so-called fundamental equation of adaptive optics, which states that at any instant $t$, the residual wavefront error $w_e$ is the difference between the incoming atmospheric turbulent wavefront $w_a$ and the mirror command\footnote{the shape of the mirror is actually set to half of the mirror command, because the OPD on a reflecting surface is twice the surface error} $w_c$ - see figure \ref{fig:01},
\begin{equation}
w_e(\br,\bth,\lambda_s,t)=
w_a(\br,\bth,\lambda_s,t)-w_c(\br,\lambda_m,t)
\label{eq:07}
\end{equation}
where \br\ is the location in the pupil plane, \bth\ is the angular separation between the science object and the guide star (assumed on-axis without loss of generality), $\lambda_s$ is the science observation wavelength, and $\lambda_m$ is the wavefront sensing wavelength. Note that in our initial paper, we used the phase instead of the wavefront in the fundamental equation, but we believe now that using the wavefront formulation is more appropriate because it is actually the wavefront which is corrected in an AO system. We will therefore develop equations for the residual wavefront s-PSD, which will be transformed into the phase s-PSD by multiplication with the usual factor $(2\pi/\lambda_s)^2$. Polychromatic PSF will be modeled by computing and averaging the phase s-PSD over the chosen optical bandwidth.

\paragraph{Including air refractivity}

The air's refractive index depends (slowly) on the wavelength, and measuring the wavefront at a different wavelength than the science observation channel introduce a small but noticeable error for systems with a tight wavefront error budget. Formally, the air's refractivity (N=n-1) is given by the sum of the refractivity of the air's constituents (nitrogen, oxygen, water, carbon dioxide etc.) Practically, though, it is shown in \cite{hill:80} that N can be written as the sum of a continuum and anomalous terms. The anomalous terms are associated with the excitation/absorption lines of water vapor and carbon dioxide (others constituents have a negligible impact), and because the atmosphere is naturally opaque at theses wavelengths, the anomalous terms are of no interest to us. The origin of the continuum term is actually not different from the anomalous terms: it is a sum of the wings of the strong nitrogen/oxygen/ozone excitation lines in the ultraviolet which extend to the visible and infrared wavelengths. There are very good theoretical/empirical models for this continuum \cite{ciddor:96,mathar:07} in the visible and infrared up to 10 $\mu m$, that are function of the air temperature, pressure, humidity and carbon dioxide content. We will not dig here into these models. What is of interest for us is that these models can be written as the product of a chromatic term which depends only on the wavelength, and a non-chromatic term which depends on the other variables (temperature, pressure etc.) Therefore, within the transmission windows of the atmosphere (i.e. inside the photometric bands), the wavefront error is simply proportional to the air's refractivity, and we can rewrite the fundamental equation of AO as
\begin{equation}
w_e(\br,\bth,\lambda_s,t)=
w_a(\br,\bth,\lambda_s,t)-\nu(\lambda_m,\lambda_s)\,w_c(\br,\lambda_s,t)
\label{eq:08}
\end{equation}
where we define $\nu(\lambda_m,\lambda_s)\equiv N(\lambda_m)/N(\lambda_s)$ as the dispersion factor. The later formulation allows us to develop the model of the mirror command $w_c$ for a single wavelength - here the science wavelength - and correct for the fact that the wavefront is actually measured at another wavelength. The effect of dispersion in discussed with more details in Jolissaint and Kendrew \cite{jolissaint:09}.

\subsection{The residual wavefront error in closed loop mode}

\paragraph{Continuous process assumption}

AO control is a discretized process: the DM shape is updated periodically at a loop period $\dt$, with a time delay $t_{\text{lag}}$ following the end of the wavefront sensor exposure. In-between these updates, the DM shape is kept constant (at least classical systems are working this way). AO control is therefore an integral and hold process, in a sense that the updated DM shape is equal to the previous one plus a weighted estimate of the wavefront residual. Now, as our approach is stationary in nature, no specific instant can be privileged: in other words, the equations we are writing need to be applicable at any instant, therefore the discrete integral control equation $c_k=g\times e_k+c_{k-1}$, where $c_k$ represents the DM command at instant $t_k$ and $g\times e_k$ the error signal weighted by the loop gain $g$, needs to be replaced by the continuous integral $c(t)=g\times\int_{0}^{t-\dt} e(t)\text{d}t+c(t-\dt)$. This approximation is equivalent to assume that the closed loop control is applied continuously, as if at any instant $t$, the DM is updated with a command computed from the WFS measurement an instant $t-t_{\text{lag}}$ earlier.

\paragraph{The deformable mirror spatial transfer function}

The command applied to the DM is made of the projection of the WFS measurement onto the DM space, whose basis is the N-dimensional set of the DM influence functions $\mathcal{I}_{i=1...N}$. The equivalence of this projection in our stationary approach is the convolution of the reconstructed wavefront with the DM spatial response, or, in the spatial frequency domain, the multiplication of the wavefront Fourier transform with the DM spatial transfer function. The DM spatial response is defined by the projection of the Dirac impulse onto the influence function basis $\mathcal{I}_{i=1...N}$,
\begin{equation}
\gamma_{\text{\tiny DM}}(\br)=\sum_{i=1}^{N}\,p_i\,\mathcal{I}_i(\br)
\label{eq:09}
\end{equation}
The coefficients $p_i$ are computed from the minimization of the quadratic distance between the Dirac impulse and its projection, and we find
\begin{gather}
\vec{p}=\mathcal{S}^{-1}\cdot\vec{b}\label{eq:10}
\\
\mathcal{S}_{i,j}=\iint\limits_{\mathbb{R}^2}\mathcal{I}_i(\br)\,\mathcal{I}_j(\br)\,\mathrm{d}^{2}r\label{eq:11}
\\
b_i=\Big\langle\iint\limits_{\mathbb{R}^2}\delta(\br-\bu)\,\mathcal{I}_i(\br)\,\mathrm{d}^{2}r\Big\rangle_{\bu\,\in\,\Box}=\big\langle\mathcal{I}_i(\bu)\big\rangle_{\bu\,\in\,\Box}\label{eq:12}
\end{gather}
where $\mathcal{S}$ is the covariance matrix of the DM influence functions.

Note that as the DM response, by definition, is not supposed to vary across the DM surface, while actually the projection of the Dirac impulse does depend on its location \bu\ within the actuators grid, we will simply consider the average DM response over all possible locations \bu. We have found that this ad-hoc procedure generates DM transfer functions that better represent transfer functions measured on real systems. Practically, as the DM actuator array is periodic, we find that $b_i$ as the average of the influence function $\mathcal{I}$ over the square space $\Box$ centered on the optical axis and of width equal to the inter-actuator pitch. For another actuator grid geometry, the space over which the DM response is averaged would be different.

\begin{figure}[htb]
   \centering
   \epsfig{file=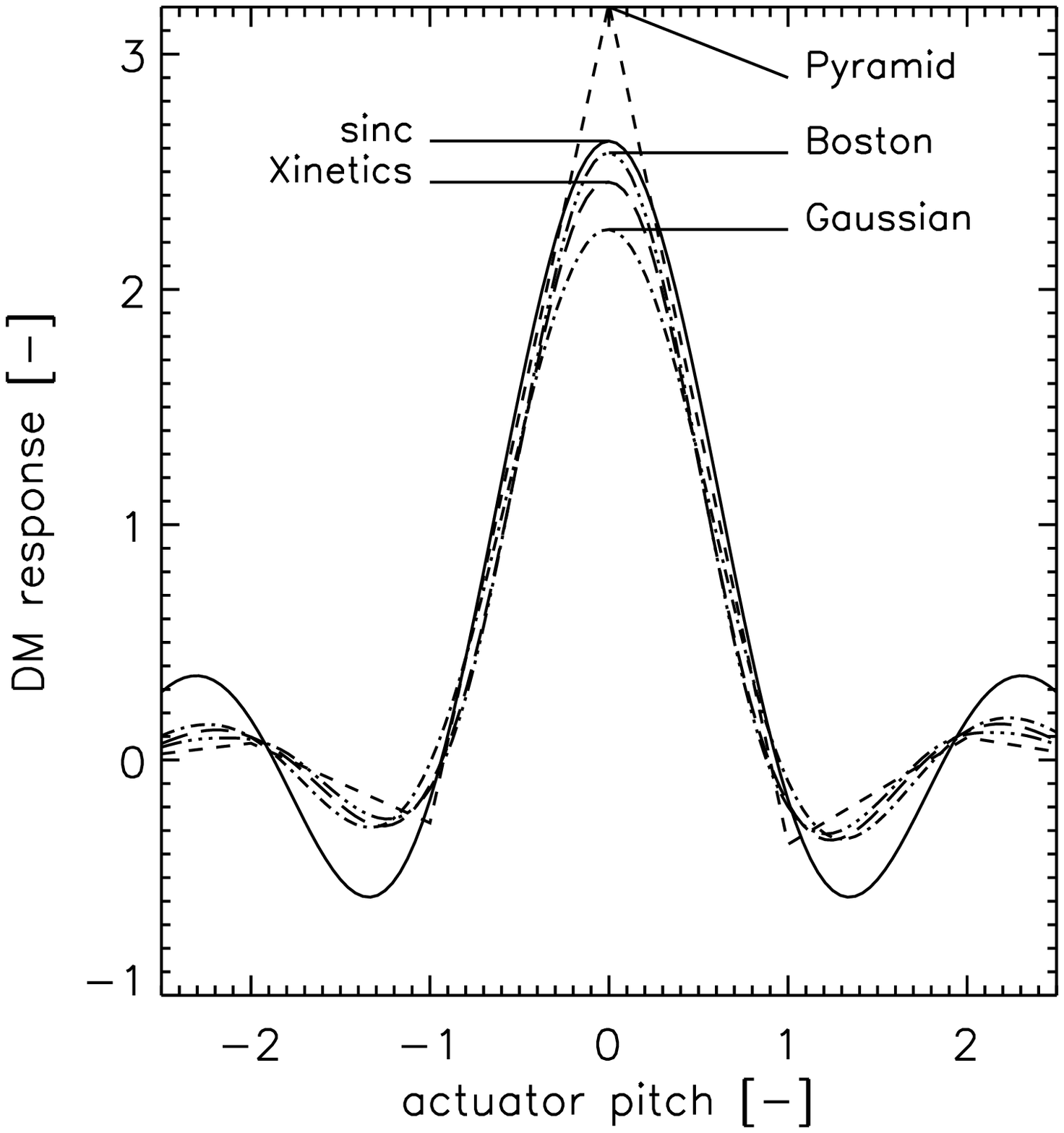,height=7.5cm,width=7.5cm}
   \hspace{5mm}
   \epsfig{file=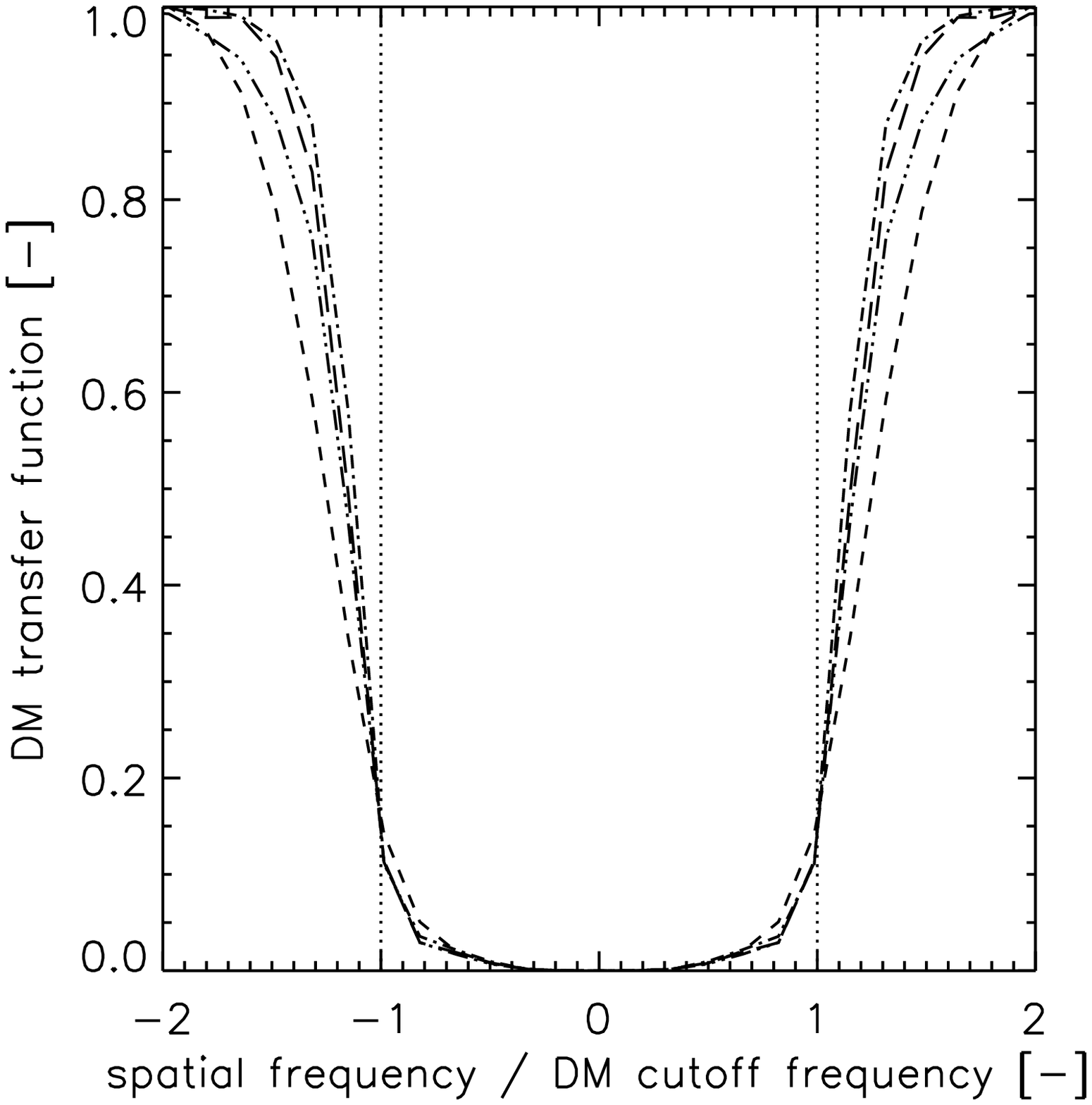,height=7.5cm,width=7.5cm}
   \caption{Left: Profile of the DM response for a Xinetics (Inc.) and a Boston MEMS deformable mirror, compared to the response function for a Gaussian and pyramid influence function model. Right: DM spatial transfer function power (modulus square) associated with the DM responses.}
   \label{fig:02}
\end{figure}

The DM spatial transfer function is given by the Fourier transform of the DM spatial response, and as the $i$-th influence function at a position $\br_{i}$ can be written as the central influence function $\mathcal{I}_0$ shifted by $-\br_{i}$, we find
\begin{equation}
\Gamma_{\text{\tiny DM}}(\bfr)=\widetilde{\mathcal{I}_0}(\bfr)\,\sum_{i=1}^{N}\,p_i\,
\exp{(-2\pi i\,\bfr\cdot\br_{i})}
\label{eq:13}
\end{equation}
where $\widetilde{\mathcal{I}_0}(\bfr)$ is the Fourier transform of the central influence function. There are numerous models for the DM influence function, all depending on the type of DM technology - see for instance \cite{huang:08}. We have developed our own empirical models for a Xinetics, Inc. 177 actuators DM model and a Boston MEMS 144 actuators model, with a modeling error of less than about 0.1 \% in amplitude, and used these to compute the DM transfer function. A cut through these two DMs response and transfer functions is shown in Figures \ref{fig:02} and is compared with Gaussian and pyramid influence function models.

It is important to note that any other DM basis of independent functions can be used to build the DM transfer function: the choice we made here of using the influence function basis was dictated by the fact that we had at our disposal accurate empirical models of several influence functions types. Now, as pointed out by a reviewer of this paper, influence functions are not always a good choice to model the DM surface, and we certainly agree that it is particularly true for low order modes (for instance tilt is badly represented by an addition of influence functions). If correction of the low order aberrations is of particular concern for the modeling of a particular system, it might therefore be worth to use a basis particularly adapted to the representation of these modes (for instance, as proposed by the reviewer, a DM surface generated by a cubic spline interpolation).

Now, let us develop the DM command $w_c$, including the continuous assumption and the DM response. The DM command $w_c$ at any instant $t$ is the update of the previous DM command (an instant $\dt$ earlier) with the residual wavefront error, weighted by the loop gain \lpg
\begin{equation}
w_c(\br,\lambda_s,t)=
\lpg\,\gdm(\br)\ast\mathcal{R}\{m(\br,\lambda_s,t)+n(\br,t)\}+
w_c(\br,\lambda_s,t-\dt)
\label{eq:14}
\end{equation}
where $\ast$ is the convolution product, $\mathcal{R}$ is the operator associated with the wavefront reconstruction from the WFS measurement $m$, and $n$ is the WFS measurement noise. Note that this equation is not specific to any type of WFS nor reconstruction operator.

The WFS output $m$ is defined by the measure (direct, gradient or laplacian) of the wavefront residual error $w_e$ averaged over the WFS integration time \dt, delayed in time by the lag $t_l$ due to the WFS readout time and the command computation. Using the notation $\overline{q}$ to indicate a time average with a time lag, we write the WFS measurement as the application of a wavefront measurement operator $\mathcal{M}$ on the instantaneous residual error $w_e$ \textbf{in the direction of the NGS}, which is assumed on-axis (i.e. \bth=0),
\begin{equation}
m(\br,\lambda_s,t)=\mathcal{M}\{\overline{w_e}(\br,0,\lambda_s,t)\}=
\mathcal{M}\{\overline{w_a}(\br,0,\lambda_s,t)-\overline{w_c}(\br,\lambda_s,t)\}
\label{eq:15}
\end{equation}

\paragraph{Separating the deformable mirror space and the wavefront analysis space}

For further developments, we need now to split the atmospheric wavefront into the component which is corrected by the DM, and the component which is simply reflected off the DM surface,
\begin{equation}
w_a=
\underbrace{w_a\ast\gamma_{\text{\tiny DM}}}_{\text{corrected by the DM}}+
\underbrace{w_a\ast(\delta-\gamma_{\text{\tiny DM}})}_{\text{reflected by the DM}}
\label{eq:16}
\end{equation}
Beside, the WFS samples the wavefront with a spatial interval $\Lambda_{\text{\tiny WFS}}$ equal to the lenslet separation distance. This naturally split the spatial frequency domain into a low spatial frequency domain, i.e. the frequencies that can be seen by the WFS, and therefore corrected, below the WFS spatial cutoff frequency
\begin{equation}
f_{\text{\tiny WFS}}=1/(2\lwfs)
\label{eq:17}
\end{equation}
and a high spatial frequency domain, above $f_{\text{\tiny WFS}}$. For a Shack-Hartmann WFS, the lenslet array has a square geometry, therefore the low spatial frequency domain is defined by the inequalities $|f_x|\le f_{\text{\tiny WFS}}$ and $|f_y|\le f_{\text{\tiny WFS}}$, a square, and the high spatial frequency domain by the complement of this square. One finally gets, with \eq{eq:16},
\begin{equation}
w_a=
(w_{a,\text{\tiny{LF}}}+w_{a,\text{\tiny{HF}}})\ast\gamma_{\text{\tiny DM}}+
(w_{a,\text{\tiny{LF}}}+w_{a,\text{\tiny{HF}}})\ast(\delta-\gamma_{\text{\tiny DM}})
\label{eq:18}
\end{equation}
The later formulation allows the separation of the DM actuators and WFS lenslet array grid architectures, which can be now studied independently.

From this point, our model development is done in the spatial frequency domain only. Our final objective is indeed to write an equivalent spatial frequency power spectrum filter for the AO correction. In the spatial frequency domain, the fundamental equation of AO becomes, where $\tilde{q}$ indicates the Fourier transform of $q$,
\begin{equation}
\widetilde{w}_e(\bfr,\bth,\lambda_s,t)=
\widetilde{w}_a(\bfr,\bth,\lambda_s,t)-
\nu(\lambda_m,\lambda_s)\,\widetilde{w}_c(\bfr,\lambda_s,t)
\label{eq:19}
\end{equation}
the DM command becomes
\begin{equation}
\widetilde{w}_c(\bfr,\lambda_s,t)=
\lpg(\bfr)\,\Gdm(\bfr)\widetilde{\mathcal{R}}
\{\widetilde{m}(\bfr,\lambda_s,t)+\tilde{n}(\bfr,t)\}+
\widetilde{w}_c(\bfr,\lambda_s,t-\dt)
\label{eq:20}
\end{equation}
and the WFS measurement
\begin{equation}
\widetilde{m}(\bfr,\lambda_s,t)=
\widetilde{\mathcal{M}}\{\overline{\widetilde{w}}_a(\bfr,0,\lambda_s,t)-
\overline{\widetilde{w}}_c(\bfr,\lambda_s,t)\}
\label{eq:21}
\end{equation}
Inserting \eq{eq:21} into \eq{eq:20}, with \eq{eq:18}, and as the reconstruction and measurement operators become spatial filters in the frequency domain, it comes (we drop momentarily the common variables to shorten the notation)
\begin{equation}
\begin{split}
\widetilde{w}_c=&
\lpg\,\Gdm(1-\Gdm)\widetilde{\mathcal{R}}\,\widetilde{\mathcal{M}}\,\overline{\widetilde{w}}_{a,\text{\tiny{LF}}}\\
&+\lpg\,\Gdm^2\widetilde{\mathcal{R}}\,\widetilde{\mathcal{M}}\,\overline{\widetilde{w}}_{a,\text{\tiny{LF}}}\\
&+\lpg\,\Gdm(1-\Gdm)\widetilde{\mathcal{R}}\,\widetilde{\mathcal{M}}\,\overline{\widetilde{w}}_{a,\text{\tiny{HF}}}\\
&+\lpg\,\Gdm^2\widetilde{\mathcal{R}}\,\widetilde{\mathcal{M}}\,\overline{\widetilde{w}}_{a,\text{\tiny{HF}}}\\
&-\lpg\,\Gdm\widetilde{\mathcal{R}}\,\widetilde{\mathcal{M}}\,\overline{\widetilde{w}}_{c}\\
&+\lpg\,\Gdm\widetilde{\mathcal{R}}\,\tilde{n}+\widetilde{w}_c(t-\dt)
\end{split}
\label{eq:22}
\end{equation}
In our approach, the product $\Gdm(1-\Gdm)$ would describe the projection onto the orthogonal of the DM space, followed by the projection onto the DM space. This product naturally has to be replaced by the null operator. Beside, the wavefront reconstruction action is to revert the WFS measurement, therefore, for a wavefront $q_{\text{\tiny{LF}}}$ strictly limited to the WFS low frequency space, the cumulated operation of wavefront measurement and wavefront reconstruction is the identity operator, i.e. $\mathcal{R}\{\mathcal{M}\{q_{\text{\tiny{LF}}}\}\}=q_{\text{\tiny{LF}}}$. With the later remarks, and noting that the DM command, by nature, belongs to the low frequency WFS space, \eq{eq:22} simplifies to
\begin{equation}
\begin{split}
\widetilde{w}_c=&
\lpg\,\Gdm^2\,\overline{\widetilde{w}}_{a,\text{\tiny{LF}}}\\
&+\lpg\,\Gdm^2\widetilde{\mathcal{R}}\,\widetilde{\mathcal{M}}\,\overline{\widetilde{w}}_{a,\text{\tiny{HF}}}\\
&-\lpg\,\Gdm\,\overline{\widetilde{w}}_{c}\\
&+\lpg\,\Gdm\widetilde{\mathcal{R}}\,\tilde{n}+
\widetilde{w}_c(t-\dt)
\end{split}
\label{eq:23}
\end{equation}

\paragraph{A note on the loop gain}

Nothing prevent us from setting the loop gain \lpg\ here as a free variable in \bfr, too. This allows optimization of the loop gain frequency-by-frequency, which is equivalent - in our stationary approach - to modal gain optimization. We will therefore keep the notation $\lpg(\bfr)$ even if loop gain optimization is not discussed further in this paper.

\paragraph{Independence of the turbulent layers and Taylor hypothesis}

In the atmosphere, optical turbulence is distributed in thin independent layers, each being characterized by (1) the so-called refractive index structure constant $C_{N,i}^2$, a measure of the variance of the refractive index spatial fluctuation within the layer, (2) the apparent velocity of the turbulent layer - see for instance \cite{azouit:05}. As seen from the pupil of the telescope, the time scale over which the wavefront associated to each layer evolves significantly is generally longer than the time it takes for the wind to push the layer across the telescope beam. Therefore, in first approximation, everything looks as if the optical turbulence profile was made of a certain number of frozen wavefront screens translating across the telescope beam with the layers wind speed and directions. This assumption of frozen optical turbulence layers is referred to as the Taylor hypothesis in the literature\footnote{departure from the Taylor hypothesis is discussed in \cite{berdja:07}}, and has the nice consequence that it is possible to transpose, within the layer, a shift in time \dt\ into a shift in space $\Delta\br=\mathbf{v}\dt$, with $\mathbf{v}$ the layer's wind velocity.

Assuming independence of the turbulent layers, the correction of the total wavefront summed over the $N_l$ turbulent layers\footnote{about 10 layers are generally needed to model a turbulent profile} is equivalent to the correction of each wavefront from each layer taken individually, as in any case, cross terms between the layers will vanish on average in the computation of the long exposure residual phase s-PSD. Let us therefore compute the wavefront error spectrum $\widetilde{w}_{e,l}$ associated with the layer $l$.

As shown in \cite{jolissaint:06}, in the Fourier domain, the time average of a wavefront $q$ over a time interval \dt, followed by a time lag \tlag\ becomes, for the layer $l$ with a wind velocity $\mathbf{v}_l$,
\begin{equation}
\overline{\tilde{q}}_l(\bfr,t)=
\sinc(\dt\,\bfr\cdot\mathbf{v}_l)
\exp{[2\pi i\,(\dt/2+\tlag)\,\bfr\cdot\mathbf{v}_l]}
\label{eq:24}
\end{equation}
where $\dt/2+\tlag$, equaling the time interval between the middle of the WFS exposure time and the application of the new DM command, represents the overall time lag. With the later, the DM command spectrum \eq{eq:23} becomes, for the layer $l$,
\begin{equation}
\begin{split}
\widetilde{w}_{c,l}=
\lpg\Gdm^2\sinc(\dt\,\bfr\cdot\mathbf{v}_l)\exp{[2\pi i\,(\dt/2+\tlag)\,\bfr\cdot\mathbf{v}_l]}\,
\widetilde{w}_{a,\text{\tiny{LF}},l}+\\
\lpg\Gdm^2\sinc(\dt\,\bfr\cdot\mathbf{v}_l)\exp{[2\pi i\,(\dt/2+\tlag)\,\bfr\cdot\mathbf{v}_l]}\,
\widetilde{\mathcal{R}}\widetilde{\mathcal{M}}\,\widetilde{w}_{a,\text{\tiny{HF}},l}-\\
\lpg\Gdm\sinc(\dt\,\bfr\cdot\mathbf{v}_l)\exp{[2\pi i\,(\dt/2+\tlag)\,\bfr\cdot\mathbf{v}_l]}\,
\widetilde{w}_{c,l}+\\
\exp{(2\pi i\,\dt\,\bfr\cdot\mathbf{v}_l)}\,\widetilde{w}_{c,l}
\end{split}
\label{eq:25}
\end{equation}
where we have replaced the time shift \dt\ within $\widetilde{w}_{c,l}(t-\dt)$ by its equivalent phase change $\exp{(2\pi i\,\dt\,\bfr\cdot\mathbf{v}_l)}$ in the Fourier domain, using the Taylor hypothesis.

The noise term, being added to the wavefront slope measurement, is not linked in any way to the turbulent layers, therefore it does not make any sense to write a noise term for each layer. The noise term is independent from the other error terms and needs to be treated separately, so we do not include any noise term in the last equation. The overall WFE will simply given by the sum of the servo-lag contribution and the noise contribution.

Regrouping the terms in $\widetilde{w}_{c,l}$, we end up with an expression for the DM command
\begin{equation}
\begin{split}
\widetilde{w}_{c,l}(\bfr,\lambda_s,t)=
\Big\{\lpg(\bfr)\,\Gdm^2(\bfr)\,\sinc(\dt\,\bfr\cdot\mathbf{v}_l)\exp{[2\pi i\,(\dt/2+\tlag)\,\bfr\cdot\mathbf{v}_l]}
\,\widetilde{w}_{a,\text{\tiny{LF}},l}(\bfr,0,\lambda_s,t)+\\
\lpg(\bfr)\,\Gdm^2(\bfr)\,\sinc(\dt\,\bfr\cdot\mathbf{v}_l)\exp{[2\pi i\,(\dt/2+\tlag)\,\bfr\cdot\mathbf{v}_l]}
\,\widetilde{\mathcal{R}}(\bfr)\,\widetilde{\mathcal{M}}(\bfr)\,\widetilde{w}_{a,\text{\tiny{HF}},l}(\bfr,0,\lambda_s,t)\Big\}\Big/\\
\Big[1+\lpg(\bfr)\,\Gdm(\bfr)\,\sinc(\dt\,\bfr\cdot\mathbf{v}_l)\exp{[2\pi i\,(\dt/2+\tlag)\,\bfr\cdot\mathbf{v}_l]}-
\exp{(2\pi i\,\dt\,\bfr\cdot\mathbf{v}_l)}\Big]
\end{split}
\label{eq:26}
\end{equation}
Inserting \eq{eq:26} into \eq{eq:19}, we get, again for the layer $l$,
\begin{multline}
\widetilde{w}_{e,l}(\bfr,\bth,\lambda_s,t)=
\widetilde{w}_{a,\text{\tiny{HF}},l}(\bfr,\bth,\lambda_s,t)+\widetilde{w}_{a,\text{\tiny{LF}},l}(\bfr,\bth,\lambda_s,t)\\
-\frac{\nu(\lambda_m,\lambda_s)
\lpg(\bfr)\,\Gdm^2(\bfr)\,\sinc(\dt\,\bfr\cdot\mathbf{v}_l)\exp{[2\pi i\,(\dt/2+\tlag)\,\bfr\cdot\mathbf{v}_l]}
\widetilde{w}_{a,\text{\tiny{LF}},l}(\bfr,0,\lambda_s,t)}
{1+\lpg(\bfr)\,\Gdm(\bfr)\,\sinc(\dt\,\bfr\cdot\mathbf{v}_l)\exp{[2\pi i\,(\dt/2+\tlag)\,\bfr\cdot\mathbf{v}_l]}-
\exp{(2\pi i\,\dt\,\bfr\cdot\mathbf{v}_l)}}\\
-\frac{\nu(\lambda_m,\lambda_s)
\lpg(\bfr)\,\Gdm^2(\bfr)\,\sinc(\dt\,\bfr\cdot\mathbf{v}_l)\exp{[2\pi i\,(\dt/2+\tlag)\,\bfr\cdot\mathbf{v}_l]}
\widetilde{\mathcal{R}}(\bfr)\widetilde{\mathcal{M}}(\bfr)\widetilde{w}_{a,\text{\tiny{HF}},l}(\bfr,0,\lambda_s,t)}
{1+\lpg(\bfr)\,\Gdm(\bfr)\,\sinc(\dt\,\bfr\cdot\mathbf{v}_l)\exp{[2\pi i\,(\dt/2+\tlag)\,\bfr\cdot\mathbf{v}_l]}-
\exp{(2\pi i\,\dt\,\bfr\cdot\mathbf{v}_l)}}
\label{eq:27}
\end{multline}

As an angular shift \bth\ is seen, at the layer altitude $h_l$, as a spatial shift $\Delta\br=h_l\bth$, using the shift theorem of the Fourier transform we get
\begin{equation}
\widetilde{w}_{a,\text{\tiny{LF}},l}(\bfr,\bth,\lambda_s,t)=
\exp{(2\pi i\,h_l\bfr\cdot\bth)}\,\widetilde{w}_{a,\text{\tiny{LF}},l}(\bfr,0,\lambda_s,t)
\label{eq:28}
\end{equation}
We can now rewrite \eq{eq:27},
\begin{equation}
\begin{split}
&\widetilde{w}_{e,l}(\bfr,\bth,\lambda_s,t)=\\
&\underbrace{\widetilde{w}_{a,\text{\tiny{HF}},l}(\bfr,\bth,\lambda_s,t)}_{\text{high order WFS "fitting" error}}\\
&+\underbrace{F_{\text{\tiny AS},l}(\bfr)\,\widetilde{w}_{a,\text{\tiny{LF}},l}(\bfr,0,\lambda_s,t)}_{\text{aniso-servo error}}\\
&+\underbrace{F_{\text{\tiny AL},l}(\bfr)\,\widetilde{w}_{a,\text{\tiny{HF}},l}(\bfr,0,\lambda_s,t)}_{\text{WFS aliasing error}}
\end{split}
\label{eq:29}
\end{equation}
and identify the four fundamental terms of the residual wavefront error:
\begin{enumerate}
\item the high order WFS error, usually named the "DM fitting error" in the AO literature - which is actually for us the part of the atmospheric wavefront which is not seen by the WFS, therefore cannot be corrected by the DM, so we think that calling this error the high order WFS error is more appropriate,
\item the angular anisoplanatic AND loop servo-lag error, identified here as the "aniso-servo" error, as anisoplanatism and servo-lag error are correlated (with the Taylor hypothesis, a wavefront shift in time can be compensated by a negative wavefront shift is space),
\item the WFS aliasing error: the high order wavefront error is seen by the WFS as a low spatial frequency error and reconstructed as such, therefore the AO system is compensating an error which actually is non-existant,
\item and finally the WFS noise term, discussed later.
\end{enumerate}
$F_{\text{\tiny AS},l}$ is defined as the aniso-servo spatial filter for the layer $l$,
\begin{multline}
F_{\text{\tiny AS},l}(\bfr)=\exp{(2\pi i\,h_l\bfr\cdot\bth)}-\\
\frac{\nu(\lambda_m,\lambda_s)
\lpg(\bfr)\,\Gdm^2(\bfr)\,\sinc(\dt\,\bfr\cdot\mathbf{v}_l)\exp{[2\pi i\,(\dt/2+\tlag)\,\bfr\cdot\mathbf{v}_l]}}
{1+\lpg(\bfr)\,\Gdm(\bfr)\,\sinc(\dt\,\bfr\cdot\mathbf{v}_l)\exp{[2\pi i\,(\dt/2+\tlag)\,\bfr\cdot\mathbf{v}_l]}-
\exp{(2\pi i\,\dt\,\bfr\cdot\mathbf{v}_l)}}
\label{eq:30}
\end{multline}
and $F_{\text{\tiny AL},l}$ is the WFS aliasing spatial filter for the layer $l$
\begin{multline}
F_{\text{\tiny AL},l}(\bfr)=\\
-\frac{\nu(\lambda_m,\lambda_s)\lpg(\bfr)\,\Gdm^2(\bfr)\,\sinc(\dt\,\bfr\cdot\mathbf{v}_l)
\exp{[2\pi i\,(\dt/2+\tlag)\,\bfr\cdot\mathbf{v}_l]}\widetilde{\mathcal{R}}(\bfr)\widetilde{\mathcal{M}}(\bfr)}
{1+\lpg(\bfr)\,\Gdm(\bfr)\,\sinc(\dt\,\bfr\cdot\mathbf{v}_l)\exp{[2\pi i\,(\dt/2+\tlag)\,\bfr\cdot\mathbf{v}_l]}-
\exp{(2\pi i\,\dt\,\bfr\cdot\mathbf{v}_l)}}
\label{eq:31}
\end{multline}
It is interesting to examine the limits of the aniso-servo and WFS aliasing spatial filters when there is no angular separation between the science object and the NGS, i.e. \bth=0, and when the WFS integration time and loop lag are set to zero, \dt=\tlag=0. We find
\begin{gather}
F_{\text{\tiny AS},l}\,\rightarrow\, 1-\nu\,\Gdm\nonumber \\
F_{\text{\tiny AL},l}\,\rightarrow\, -\nu\,\Gdm\,\widetilde{\mathcal{R}}\,\widetilde{\mathcal{M}}\nonumber
\end{gather}
which indicates that in the absence of aniso-servo error, the residual low frequency wavefront error is generated by (1) the refractive index dispersion -- and we see as we would expect that this error is proportional to the dispersion factor, and (2) the aberrations seen by the WFS that the DM cannot correct, which are never null because even if the DM actuator pitch is equal to the WFS lenslet pitch, a perfect correction of the low spatial frequencies would require a DM with sinus cardinal influence function (Fourier transform of a sinus cardinal is a door function), which is only approximated by actual influence functions - see Figure \ref{fig:02}. The same is true for the WFS aliasing error. This behavior corresponds well to what we would have expected.

\paragraph{A note on the correlation between the error terms} Computing the s-PSD associated with the four fundamental wavefront error terms above essentially consists in computing the modulus square of \eq{eq:29}, averaged over the time $t$. Therefore, cross products appear between the four error terms. Now, the noise term is naturally not correlated with the other errors and can be treated separately. The high order WFS error is not seen by the system and transmitted to the output of the system, unaffected. We will assume in this paper, without discussing it further, that the cross products between the low and high order spatial frequencies are negligible relative to the main error terms. Therefore, in what follows, the four terms above will be discussed independently from each others.

We will now make use of the expressions developed above for the \textit{wavefront} error terms to develop the analytical expressions of the residual \textit{phase} s-PSD of the four fundamental errors we have identified. As the residual phase s-PSD is given by spatial filtering of the optical turbulence phase s-PSD, we will start by recalling the expression of the later, as given in the literature.

\subsection{The turbulent phase spatial power spectrum}

The s-PSD of the turbulent phase is discussed in Roddier \cite{roddier:81}. In the atmosphere, the extension of optical turbulence is necessarily limited by the individual layers thickness, and an optical turbulence outer scale $L_0$ was included in the Kolmogorov s-PSD to account for this spatial limitation. This modified Kolmogorov s-PSD is called the von Karm\`an s-PSD in the literature (see for instance Winker \cite{winker:91}, and Maire et al. \cite{maire:08} for a few other s-PSD models) and is given, at the wavelength $\lambda$, by
\begin{equation}
\Xi_{\text{\tiny ATM}}(\bfr)=
0.0229\,r_0(\lambda)^{-5/3}(|\bfr|^2+1/L_0^2)^{-11/6}
\label{eq:32}
\end{equation}
where $r_0$ is the Fried parameter, a measure of the strength of turbulence, defined as the telescope diameter whose focal plane angular frequency cutoff would be the same than the optical turbulence cutoff frequency (see Fried \cite{fried:66}). $r_0$ is generally given at 500 nm in the literature, and we will follow this convention, unless indicated differently. Typical values for $r_0$ at 500 nm extend from 5 cm (bad observation site, day-light conditions) to 25 cm (excellent site). The optical outer scale is generally in the range 20 to 40 m, surprisingly with very few variations between the different sites where this quantity has been measured. 

The Fried parameter is associated to the vertical profile of the optical turbulence structure constant $C_N^2(h)$, following
\begin{equation}
r_0(\lambda)^{-5/3}=
0.4234\,(2\pi/\lambda)^2\int_{0}^{\infty}\,C_N^2(h)\,\text{d}h
\label{eq:33}
\end{equation}
which can be written, in the case of $N_l$ independent optical turbulence layers, as a sum
\begin{equation}
r_0(\lambda)^{-5/3}=
0.4234\,(2\pi/\lambda)^2\sum_{l=1}^{N_l}\,C_{N,l}^2\,\Delta h_l
=\sum_{l=1}^{N_l}\,r_{0,l}(\lambda)^{-5/3}
\label{eq:34}
\end{equation}
where $\Delta h_l$ is the layer's thickness, and $r_{0,l}$ defines the layer's Fried parameter. Consequently we can also define a phase s-PSD for each layer,
\begin{equation}
\Xi_{\text{\tiny ATM},l}(\bfr)=
0.0229\,r_{0,l}(\lambda)^{-5/3}(|\bfr|^2+1/L_0^2)^{-11/6}
\label{eq:35}
\end{equation}
which naturally sums up to the overall phase s-PSD,
\begin{equation}
\Xi_{\text{\tiny ATM}}(\bfr)=\sum_{l=1}^{N_l}\,\Xi_{\text{\tiny ATM},l}(\bfr)
\label{eq:36}
\end{equation}
Using this notation, the phase s-PSD associated with the low and high order turbulent wavefront errors - $w_{a,\text{\tiny{LF}},l}$ and $w_{a,\text{\tiny{HF}},l}$ - will be written now
\begin{gather}
(2\pi/\lambda_s)^2\langle|\widetilde{w}_{a,\text{\tiny{LF}},l}(\bfr,0,\lambda_s,t)|^2\rangle_t=
\maskl(\bfr)\,\Xi_{\text{\tiny ATM},l}(\bfr)\label{eq:37}\\
(2\pi/\lambda_s)^2\langle|\widetilde{w}_{a,\text{\tiny{HF}},l}(\bfr,0,\lambda_s,t)|^2\rangle_t=
\maskh(\bfr)\,\Xi_{\text{\tiny ATM},l}(\bfr)\label{eq:38}
\end{gather}
where \maskl\ and \maskh\ are low and high spatial frequency masks, defined, for a Shack-Hartmann WFS, by the square domain
\begin{gather}
\maskl(\bfr)=
\begin{cases}
1 & |f_x|,|f_y|\le f_{\text{\tiny WFS}}\\
0 & \text{elsewhere}
\end{cases}\label{eq:39}
\\
\text{and}\ \ \maskh(\bfr)=1-\maskl(\bfr)\label{eq:40}
\end{gather}

\subsection{The high order WFS spatial power spectrum - or "fitting error"}

The high order WFS phase error s-PSD is simply given by the atmospheric turbulence phase s-PSD limited to the high spatial frequency domain and is written
\begin{equation}
\Xi_{\text{\tiny HF}}(\bfr)=\maskh(\bfr)\,\Xi_{\text{\tiny ATM}}(\bfr)
\label{eq:41}
\end{equation}
where $\Xi_{\text{\tiny ATM}}$ is given by \eq{eq:32}.

\subsection{The aniso-servo spatial power spectrum}

The aniso-servo phase error s-PSD is given by the time average of the modulus square of the aniso-servo wavefront error, translated into a phase error, summed over the $N_l$ turbulent layers. From \eq{eq:29} and \eq{eq:37}, we get
\begin{equation}
\Xi_{\text{\tiny AS}}(\bfr)=
\maskl(\bfr)\,\sum_{l=1}^{N_l}\,|F_{\text{\tiny AS},l}|^2(\bfr)\,
\Xi_{\text{\tiny ATM},l}(\bfr)
\label{eq:42}
\end{equation}
where $\Xi_{\text{\tiny ATM},l}$ is given in \eq{eq:35} and
\begin{multline}
|F_{\text{\tiny AS},l}|^2(\bfr)=\Big(1+\lpg^2(\bfr)\Gdm^2(\bfr)\,\sinc^2(\dt\,\bfr\cdot\mathbf{v}_l)
[1+\nu^2(\lambda_m,\lambda_s)\Gdm^2(\bfr)]/2-\cos{(2\pi\dt\bfr\cdot\mathbf{v}_l)}\\
+\lpg(\bfr)\Gdm^2(\bfr)\,\sinc(\dt\,\bfr\cdot\mathbf{v}_l)\nu(\lambda_m,\lambda_s)\times\\
\big\{\cos{[2\pi h_l\bfr\cdot\bth+2\pi(\dt/2-\tlag)\bfr\cdot\mathbf{v}_l]}-
  \cos{[2\pi h_l\bfr\cdot\bth-2\pi(\dt/2+\tlag)\bfr\cdot\mathbf{v}_l]}\big\}\\
+\lpg(\bfr)\Gdm(\bfr)\,\sinc(\dt\,\bfr\cdot\mathbf{v}_l)
\big\{\cos{[2\pi(\dt/2+\tlag)\bfr\cdot\mathbf{v}_l]}-\cos{[2\pi(\dt/2-\tlag)\bfr\cdot\mathbf{v}_l]}\big\}\\
-\lpg^2(\bfr)\Gdm^3(\bfr)\,\sinc^2(\dt\,\bfr\cdot\mathbf{v}_l)\nu(\lambda_m,\lambda_s)\cos{(\pi h_l\bfr\cdot\bth)}\Big)\Big/\\
\Big(1+\lpg^2(\bfr)\Gdm^2(\bfr)\,\sinc^2(\dt\,\bfr\cdot\mathbf{v}_l)/2+\lpg(\bfr)\Gdm(\bfr)\,\sinc(\dt\,\bfr\cdot\mathbf{v}_l)\times\\
\big\{\cos{[2\pi(\dt/2+\tlag)\bfr\cdot\mathbf{v}_l]}-\cos{[2\pi(\dt/2-\tlag)\bfr\cdot\mathbf{v}_l]}\big\}-
\cos{(2\pi\dt\bfr\cdot\mathbf{v}_l)}\Big)
\label{eq:43}
\end{multline}
While this equation seems impressive, coding it into a computer program does not represent a particular challenge.

\subsection{The WFS aliasing spatial power spectrum}

The aliasing error is given, for the layer $l$, by the term - see \eq{eq:27}
\begin{multline}
\widetilde{w}_{\text{\tiny AL},l}(\bfr,t)=
-\widetilde{\mathcal{R}}(\bfr)\widetilde{\mathcal{M}}(\bfr)\widetilde{w}_{a,\text{\tiny{HF}},l}(\bfr,0,\lambda_s,t)\times
\\
\frac{\nu(\lambda_m,\lambda_s)
\lpg(\bfr)\,\Gdm^2(\bfr)\,\sinc(\dt\,\bfr\cdot\mathbf{v}_l)\exp{[2\pi i\,(\dt/2+\tlag)\,\bfr\cdot\mathbf{v}_l]}}
{1+\lpg(\bfr)\,\Gdm(\bfr)\,\sinc(\dt\,\bfr\cdot\mathbf{v}_l)\exp{[2\pi i\,(\dt/2+\tlag)\,\bfr\cdot\mathbf{v}_l]}-
\exp{(2\pi i\,\dt\,\bfr\cdot\mathbf{v}_l)}}
\label{eq:44}
\end{multline}
the product $\widetilde{\mathcal{R}}\,\widetilde{\mathcal{M}}\,\widetilde{w}_{a,\text{\tiny{HF}},l}$ was already developed in our initial paper, and is recalled here. A Shack-Hartmann WFS produces a measurement of the wavefront slope (gradient) in both x and y directions, with a spatial sampling given by the WFS lenslet array pitch \lwfs\ - which is also the lenslet width. In the spatial frequency domain, the slope measurement is given by the multiplication of the wavefront Fourier transform with the two components operator (one for each direction)

\begin{multline}
\widetilde{\mathcal{M}}(\bfr)=
[\widetilde{\mathcal{M}}_x(\bfr),\widetilde{\mathcal{M}}_y(\bfr)]=
2\pi i\,\lwfs^{2}\,[\bfr\,\sinc(\lwfs\,f_x)\,\sinc(\lwfs\,f_y)]
\ast\comb(\lwfs\bfr)
\label{eq:45}
\end{multline}
where the product with the spatial frequency vector \bfr\ stands for the derivative in the Fourier domain, the \sinc\ function is for the wavefront average over the lenslet area, and $\comb(\lwfs\bfr)$, the Dirac comb, represents the recurrence of the measured slope spectrum with a spacing 2\fwfs\ in both x and y directions, and is responsible for the aliasing of the part of the spectrum above the WFS cutoff frequency \fwfs\ inside the low spatial frequency domain -- which always occurs, because the turbulent wavefront spectrum is not band limited at high spatial frequency.

The analytical expression for the reconstruction operator Fourier transform $\widetilde{\mathcal{R}}$ is computed from the minimization of the quadratic distance between the slope measurement and the actual slope (least square error algorithm, LSE). The weakness of the LSE algorithm is that the WFS noise is reconstructed as a real signal, without penalty. Other algorithms have therefore been proposed that make use of a priori knowledge of the Kolmogorov-statistics based signal and noise statistics to minimize the contribution of the noise on the reconstructed signal - see for instance \cite{neichel:08}. A discussion of the pros and cons of these different algorithms is beyond the scope of this paper, though, so we will stick with the LSE-based algorithm, as it is the most simple and most straightforward to implement.

As we have seen, the slope measurement operator in the Fourier domain is, basically, a multiplication with the spatial frequency vector \bfr. The reconstruction operator is therefore the inverse operator, i.e. the inverse of the vector\footnote{$\ \vec{u}\cdot\vec{v}=1$ has the solution $\vec{u}=\vec{v}/|\vec{v}|^2$} \bfr, but ignoring the Dirac comb, because the reconstruction does not extend beyond the WFS cutoff frequency \fwfs, and we find (the factor \lwfs\ disappears because it is actually part of the Dirac comb convolution product)
\begin{equation}
\widetilde{\mathcal{R}}(\bfr)=[\widetilde{\mathcal{R}}_x(\bfr),\widetilde{\mathcal{R}}_y(\bfr)]=
\frac{\bfr}{2\pi i\,|\bfr|^{2}\,\sinc(\lwfs\,f_x)\,\sinc(\lwfs\,f_y)}
\label{eq:46}
\end{equation}
We can now develop the term $\widetilde{\mathcal{R}}\,\widetilde{\mathcal{M}}\,\widetilde{w}_{a,\text{\tiny{HF}},l}$ from \eq{eq:44}. Using the two equalities
\begin{equation}
\comb(\lwfs\bfr)=\frac{1}{\lwfs^2}
\sum_{m,n=-\infty}^{\infty}\,\delta(f_x-\frac{m}{\lwfs},f_y-\frac{n}{\lwfs})
\label{eq:47}
\end{equation}
and
\begin{equation}
\frac{\sinc(\lwfs f_{[x,y]}-[m,n])}{\sinc(\lwfs f_{[x,y]})}=
\frac{(-1)^{[m,n]}\lwfs f_{[x,y]}}{\lwfs f_{[x,y]}-[m,n]}
\label{eq:48}
\end{equation}
we find
\begin{multline}
\widetilde{\mathcal{R}}(\bfr)\widetilde{\mathcal{M}}(\bfr)\widetilde{w}_{a,\text{\tiny{HF}},l}(\bfr,0,\lambda_s,t)=
\frac{f_x\,f_y}{|\bfr|^2}\times\\
\sum_{\substack{m,n=-\infty\\
|m|+|n|> 0}}^{\infty}\,(-1)^{m+n}\,\left(\frac{f_x}{f_y-\frac{n}{\lwfs}}+\frac{f_y}{f_x-\frac{m}{\lwfs}}\right)
\widetilde{w}_{a,l}(f_x-\frac{m}{\lwfs},f_y-\frac{n}{\lwfs},0,\lambda_s,t)
\label{eq:49}
\end{multline}
where it is important to note that $|m|+|n|>0$ because we do not want to include the low frequency part of the wavefront spectrum in the sum (as it is of course not aliased).

We can give now the expression for the WFS aliasing \textit{phase} error s-PSD. Summed over the $N_l$ independent layers, it is given by
\begin{equation}
\Xi_{\text{\tiny AL}}(\bfr)=\maskl(\bfr)\,\sum_{l=1}^{\infty}\,\Xi_{\text{\tiny AL},l}(\bfr)
\label{eq:50}
\end{equation}
where, for each layer,
\begin{multline}
\Xi_{\text{\tiny AL},l}(\bfr)=\langle|\widetilde{w}_{\text{\tiny AL},l}(\bfr,t)|^2\rangle_t=
\nu^2(\lambda_m,\lambda_s)\,\lpg^2(\bfr)\,\Gdm^4(\bfr)\,\sinc^2(\dt\,\bfr\cdot\mathbf{v}_l)\Big/\\
\Big(1+\lpg^2(\bfr)\Gdm^2(\bfr)\,\sinc^2(\dt\,\bfr\cdot\mathbf{v}_l)/2+\lpg(\bfr)\Gdm(\bfr)\,\sinc(\dt\,\bfr\cdot\mathbf{v}_l)\times\\
\{\cos{[2\pi(\dt/2+\tlag)\bfr\cdot\mathbf{v}_l]}-\cos{[2\pi(\dt/2-\tlag)\bfr\cdot\mathbf{v}_l]}\}-
\cos{(2\pi\dt\bfr\cdot\mathbf{v}_l)}\Big)\times\\
\frac{f_x^2\,f_y^2}{|\bfr|^4}\sum_{\substack{m,n=-\infty\\|m|+|n|> 0}}^{\infty}\,
\left(\frac{f_x}{f_y-\frac{n}{\lwfs}}+\frac{f_y}{f_x-\frac{m}{\lwfs}}\right)^2\,
\Xi_{\text{\tiny ATM},l}(f_x-\frac{m}{\lwfs},f_y-\frac{n}{\lwfs})
\label{eq:51}
\end{multline}
where we made the assumption that the correlation of the phase for frequencies separated by a 2\fwfs\ interval is negligible. It is important to realize that the term 
$T=\frac{f_x^2\,f_y^2}{|\bfr|^4}\sum(...)\Xi_{\text{\tiny ATM},l}(...)$ in \eq{eq:51} has singularities at $(f_x=0,m=0)$, $(f_y=0,n=0)$ and \bfr=0. Computation of the limits gives
\begin{equation}
\begin{cases}
T(0,f_y)=\sum_{\substack{n=-\infty\\|n|> 0}}^{\infty}\,\Xi_{\text{\tiny ATM},l}(0,f_y-\frac{n}{\lwfs})\\
T(f_x,0)=\sum_{\substack{m=-\infty\\|m|> 0}}^{\infty}\,\Xi_{\text{\tiny ATM},l}(f_x-\frac{m}{\lwfs},0)
\label{eq:52}
\end{cases}
\end{equation}

\subsection{The WFS noise spatial power spectrum}

From \eq{eq:23}, it comes
\begin{equation}
\Xi_{\text{\tiny NS}}(\bfr)=\langle|\widetilde{w}_{\text{\tiny NS}}(\bfr,t)|^{2}\rangle=
\nu^{2}(\lambda_m,\lambda_s)\,\lpg^{2}(\bfr)\,\Gdm^{2}(\bfr)\,
\langle|\widetilde{\mathcal{R}}(\bfr)\,\tilde{n}(\bfr,t)|^{2}\rangle
\label{eq:53}
\end{equation}
What is the noise $n(\br,t)$ made of, exactly ? it is a discrete quantity, in space and time, made of two components in x and y, $n(\br,t)=[n_x(\br,t),n_y(\br,t)]$ sampled on a grid of spacing \lwfs. Its spatial spectrum is therefore necessarily limited to the domain $|f_x|,|f_y|\le1/(2\lwfs)$. As the noise over the lenslets is uncorrelated, all values are possible at any instant and location, therefore the noise s-PSD is necessarily white, i.e. it is constant in the domain $|f_x|,|f_y|\le1/(2\lwfs)$, and is the same for both x and y components. So, we can define
\begin{equation}
\langle|\tilde{n}_x(\bfr,t)|^2\rangle_t = 
\langle|\tilde{n}_y(\bfr,t)|^2\rangle_t =
\widetilde{\mathcal{N}}^2(\bfr) = \text{constant}
\label{eq:54}
\end{equation}
such that
\begin{equation}
\iint\limits_{\substack{|f_x|,|f_y|\le\\1/(2\lwfs)}}\widetilde{\mathcal{N}}^2(\bfr)\text{d}^2f=
\sigma_{\text{\tiny NEA, CL}}^2=
\widetilde{\mathcal{N}}^2/\lwfs^2
\label{eq:55}
\end{equation}
so,
\begin{equation}
\widetilde{\mathcal{N}}^2=\lwfs^2\,\sigma_{\text{\tiny NEA, CL}}^2
\label{eq:56}
\end{equation}
where $\sigma_{\text{\tiny NEA, CL}}^2$ is the closed loop noise equivalent angle (NEA) variance, discussed later. The noise s-PSD is given by the average modulus square of $\widetilde{w}_{\text{\tiny NS}}$. With the reconstructor Fourier transform -- given in \eq{eq:46}, it comes
\begin{multline}
\langle|\widetilde{\mathcal{R}}(\bfr)\,\tilde{n}(\bfr,t)|^2\rangle=
\langle|\widetilde{\mathcal{R}}_x(\bfr)\,\tilde{n}_x(\bfr,t)|^2\rangle+
\langle|\widetilde{\mathcal{R}}_y(\bfr)\,\tilde{n}_y(\bfr,t)|^2\rangle\\
=\frac{f_x^2\langle|\tilde{n}_x(\bfr,t)|^2\rangle+
       f_y^2\langle|\tilde{n}_y(\bfr,t)|^2\rangle}
       {4\pi^2\,|\bfr|^{4}\,\sinc^2(\lwfs\,f_x)\,\sinc^2(\lwfs\,f_y)}
=\frac{\lwfs^2\,\sigma_{\text{\tiny NEA, CL}}^2}
{4\pi^2\,|\bfr|^{2}\,\sinc^2(\lwfs\,f_x)\,\sinc^2(\lwfs\,f_y)}
\label{eq:57}
\end{multline}
therefore we get, from \eq{eq:53},
\begin{equation}
\Xi_{\text{\tiny NS}}(\bfr)=
\frac{\nu^2(\lambda_m,\lambda_s)\,\Gdm^2(\bfr)\,\lwfs^2\,\sigma_{\text{\tiny NEA, CL}}^2}
{4\pi^2\,|\bfr|^{2}\,\sinc^2(\lwfs\,f_x)\,\sinc^2(\lwfs\,f_y)}
\label{eq:58}
\end{equation}
Note that the loop gain does not appear anymore directly in the noise s-PSD. Indeed, the noise s-PSD has to be seen as a spatial filter, actually not different from its formulation in open loop, but where the NEA is now a closed loop NEA. In other words, it is the NEA which is affected by the closed loop noise transfer function, not the spatial properties of the noise s-PSD. Note also that there is no analogy between the spatial frequency white noise and this open loop white noise. Indeed, any type of temporal spectrum is possible for the NEA signal on the lenslets, and as the lenslet noise is decorrelated from a lenslet to another, all noise distribution have the same probability over the lenslet array, therefore the spatial noise distribution is white whatever the lenslet noise statistics. In other words, it is the independence of the noise from a lenslet to another which enable the separation of the noise s-PSD from the noise temporal PSD.

We discuss now the closed loop NEA variance. Let us consider a classical model of the loop architecture: a WFS with an integration time \dt, followed by a delay \tlag\ due to the WFS readout time and the command computation time, then a numerical integral controller with gain \lpg, a digital to analog converter, and the DM. The noise rejection temporal transfer function, defined as the ratio between the NEA signal and the residual error, is given, in the steady state, by (see Demerle et al. \cite{demerle:94} for a detailed discussion),
\begin{equation}
H_n(\nu)=-\frac{\lpg(\bfr)\exp{(-2\pi i\tlag\,\nu)}\,2\pi i\dt\,\nu}
{-(2\pi\dt\,\nu)^2+\lpg(\bfr)\exp{(-2\pi i\tlag\,\nu)}[1-\exp{(-2\pi i\dt\,\nu)}]}
\label{eq:59}
\end{equation}
where $\nu$ is the temporal frequency. The noise power transfer function is given by the modulus square of $H_n$, and we find
\begin{multline}
|H_n(\nu)|^2=\\
\frac{\lpg(\bfr)^2\,(2\pi\dt\,\nu)^2}
{(2\pi\dt\,\nu)^4\!-\!2\,\lpg(\bfr)(2\pi\dt\,\nu)^2\{\cos{(2\pi\tlag\,\nu)}\!-\!\cos{[2\pi(\dt\!+\!\tlag)\,\nu]}\}
\!+\!2\,\lpg^2[1\!-\!\cos{(2\pi\dt\,\nu)}]}
\label{eq:60}
\end{multline}
The closed loop NEA variance $\sigma^2_{\text{\tiny NEA, CL}}$ is given by the integral of the filtered open loop temporal power spectrum, and as the later is a white noise limited to the domain $|\nu|<1/(2\dt)$, we find
\begin{equation}
\sigma^2_{\text{\tiny NEA, CL}}=2\,\dt\,\sigma^2_{\text{\tiny NEA, OL}}
\int\limits_{0}^{1/(2\dt)}\,|H_n(\nu)|^2\,\text{d}\nu
\label{eq:61}
\end{equation}
The open loop NEA variance $\sigma^2_{\text{\tiny NEA, OL}}$ depends on the number of NGS photons received per lenslets during the WFS integration time, the WFS geometry (lenslet width), the WFS integration time, the WFS detector read noise, and the NGS image size, which is tilt-compensated for a closed loop system. Several models have been developed in the literature for this term, and will not be reproduced here - see for instance Rousset \cite{rousset:94} and Thomas et al. \cite{thomas:06}.

Our closed loop phase s-PSD model is now complete, and is given by the sum of the s-PSD of the four fundamental AO errors: the high frequency WFS error, the aniso-servo error, the WFS aliasing error, and the WFS noise error,
\begin{equation}
\Xi_{\varphi}(\bfr)=
\Xi_{\text{\tiny HF}}(\bfr)+\Xi_{\text{\tiny AS}}(\bfr)+
\Xi_{\text{\tiny AL}}(\bfr)+\Xi_{\text{\tiny NS}}(\bfr)
\label{eq:62}
\end{equation}
Let us now illustrate the usefulness and usage of the synthetic model with a few examples.

\begin{table}[p]
\caption{Optical turbulence profile used in our illustrative examples. Paranal observatory type (taken from "E-ELT AO design inputs: relevant
atmospheric parameters", ESO document E-SPE-ESO-276-0206, except for the wind direction, which is set arbitrarily)}
\begin{center}
\begin{tabular}{cccc}
height & $C_N^2\Delta h$ & wind & wind \\
above & distr. & speed & dir.\\
pupil m & \% & $\mathrm{m\,s}^{-1}$ & /x-axis \\
\hline
42 & 53.28 & 15 & $38^{\circ}$ \\
140 & 1.45 & 13 & $34^{\circ}$ \\
281 & 3.5 & 13 & $54^{\circ}$ \\
562 & 9.57 & 9 & $42^{\circ}$ \\
1125 & 10.83 & 9 & $57^{\circ}$ \\
2250 & 4.37 & 15 & $48^{\circ}$ \\
4500 & 6.58 & 25 & $-102^{\circ}$ \\
9000 & 3.71 & 40 & $-83^{\circ}$ \\
18000 & 6.71 & 21 & $-77^{\circ}$\\
\hline
\end{tabular}
\end{center}
\label{tab:01}
\caption{Synthetic model parameters values used in our illustrative examples. Optical turbulence parameters are given at 500 nm.}
\begin{center}
\begin{tabular}{r|l}
telescope diameter & D=8 m \\
seeing angle & $w_0=0.65"$  \\
Fried parameter & $r_0=15.5$ cm \\
outer scale & $L_0=25$ m \\
phase time scale & $\tau_0=3$ ms \\
isoplanatic angle & $\theta_0=2.4"$ \\
dispersion factor & $\nu=0.99$ \\
DM conjugation & to pupil \\
DM pitch & -- free -- \\
DM actuator geometry & square \\
DM influence function & Xinetics, Inc.
\end{tabular}
\begin{tabular}{r|l}
WFS type & SH \\
WFS lenslet width & -- free -- \\
WFS throughput & 31\% \\
WFS detector noise & 2 e/px \\
WFS integration time & -- free -- \\
loop time lag & 0.8 ms \\
loop gain & -- free -- \\
NGS location & -- free -- \\
NGS magnitude & -- free -- \\
NGS BB temp & 5700 K \\
\end{tabular}
\end{center}
\label{tab:02}
\caption{wavefront errors RMS for the example discussed in section \ref{sec:41}}
\begin{center}
\begin{tabular}{r|l}
high frequency & 47 nm\\
WFS aliasing & 25 nm\\
aniso-servo & 96 nm\\
WFS noise & 142 nm\\
\hline
total error & 179 nm
\end{tabular}
\end{center}
\label{tab:03}
\end{table}

\newpage

\section{ILLUSTRATIVE EXAMPLES}

The synthetic method has been coded into our AO modeling code PAOLA\footnote{Performance of Adaptive Optics for Large (or Little) Apertures}, a general purpose IDL-based toolbox for modeling the AO correction of segmented telescope static and optical turbulence aberrations. It includes open and closed loop single NGS mode, and a complete multiple NGS ground layer AO mode. We present in this section several studies undertaken with PAOLA, to illustrate the usefulness and usage of such a synthetic tool. An optical turbulence profile and standard telescope and AO system parameters are defined in the Tables \ref{tab:01} and \ref{tab:02} to be used in the different examples.

\subsection{Structure of the residual phase spatial power spectrum, and its impact on the PSF wings}
\label{sec:41}

We consider in this example a DM with a square actuator grid, an actuator pitch of 20 cm (as projected in the telescope primary mirror), a SH-WFS lenslet width of 20 cm (as projected in M1), a WFS integration time of 2 ms, a loop gain of 0.5, an off-axis NGS at 3" and a NGS magnitude mV=12 (other parameters are given in the Tables \ref{tab:01} and \ref{tab:02}). The wavefront RMS error for the four classical components for this example are given in Table \ref{tab:03}.

\begin{figure}[htb]
   \centering
   \epsfig{file=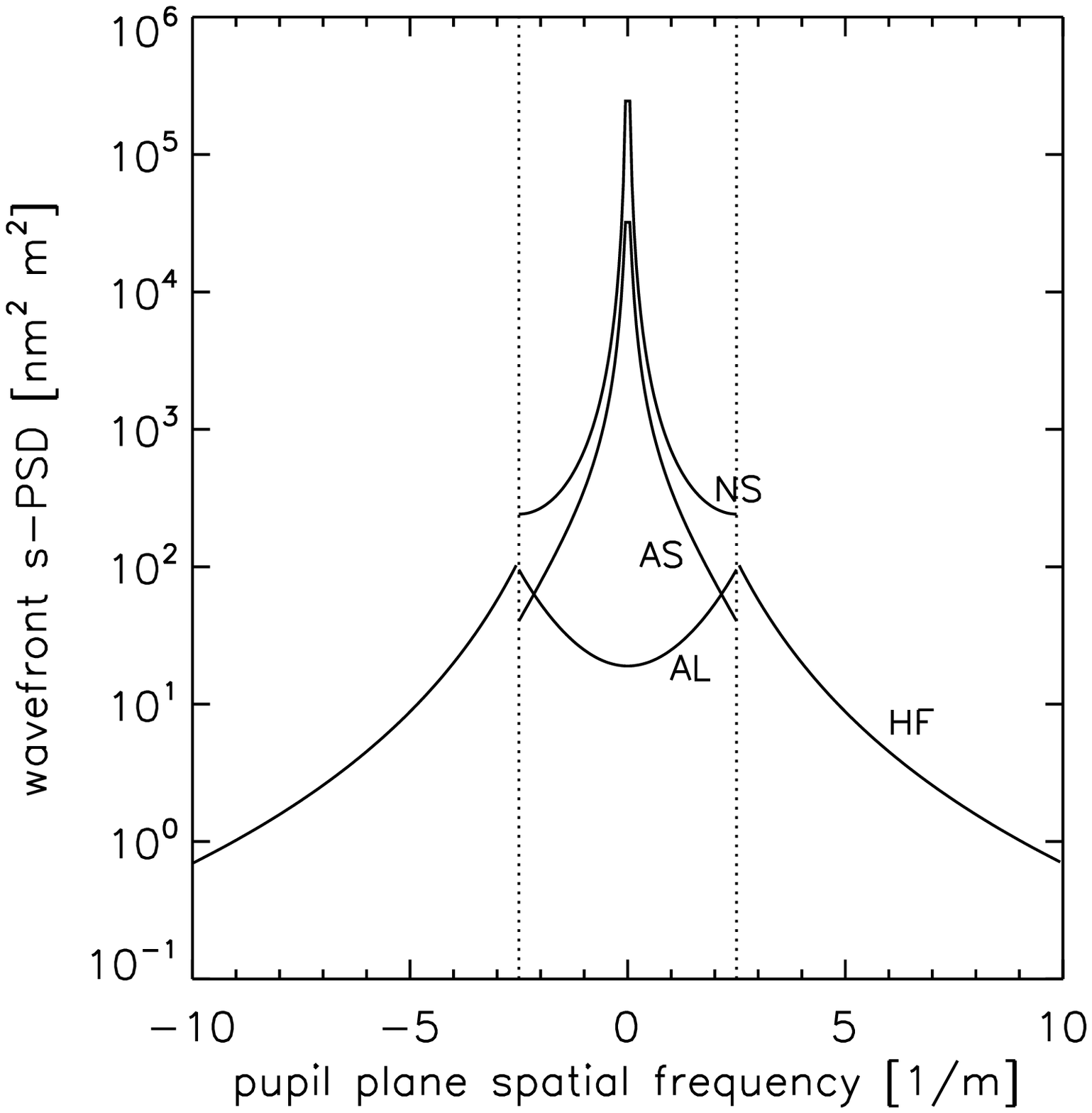,height=7.7cm,width=7.7cm}
   \hspace{2mm}
   \epsfig{file=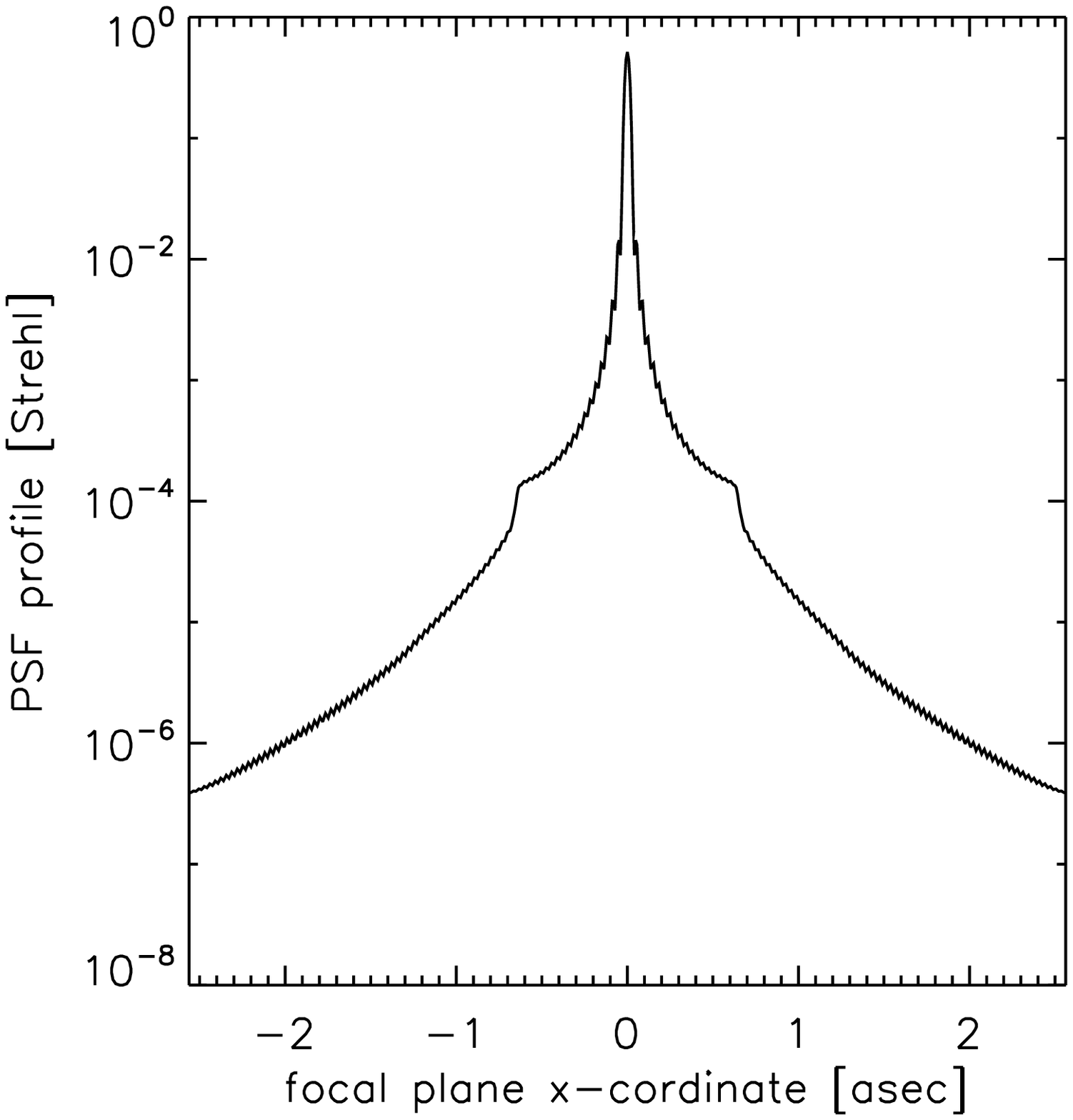,height=7.7cm,width=7.7cm}
   \caption{Profile of the four fundamental wavefront errors s-PSD, for the case discussed in section \ref{sec:41}. HF is for high frequency error, AL for WFS aliasing, AS for aniso-servo and NS is for WFS noise. The vertical dotted lines show the transition low/high spatial frequencies, at 2.5 $\text{m}^{-1}$. Right: Profile of the PSF associated with the s-PSD, at 1.25 $\mu\text{m}$. The spatial frequency and the angular coordinate are at the same scale in both figures, i.e. $x=\lambda f$. The transition core to halo occurs at an AO radius of $0.64"$.}h
   \label{fig:04}
\end{figure}
\begin{figure}[h]
   \centering
   \epsfig{file=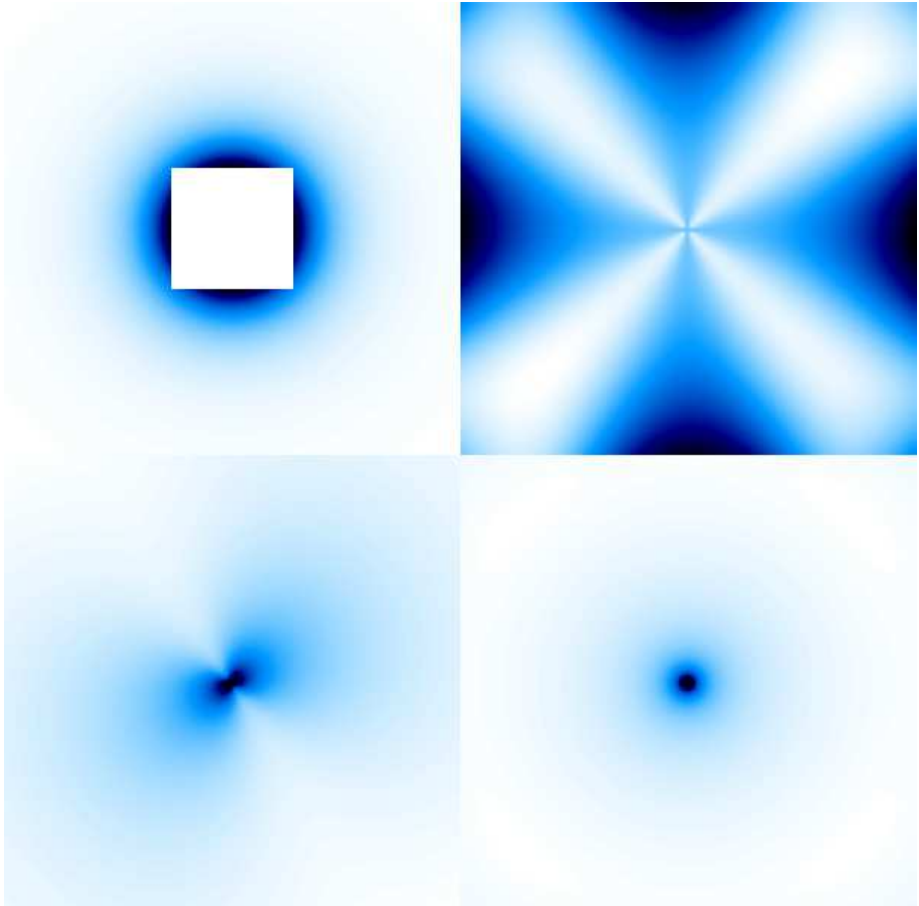,height=12cm,width=12cm}
   \caption{s-PSD of the four fundamental wavefront errors, for the case discussed in section \ref{sec:41}. Top-left: high frequency error. The central black square shows the low spatial frequency domain inside $\pm\fwfs$ (here equal to $\pm2.5\ \text{m}^{-1}$). Top-right: WFS aliasing error. Bottom-left: aniso-servo error; the s-PSD is elongated in a direction which is a composition of the main wind direction ($+54^{\circ}$ relative to the horizontal axis) and the NGS orientation (along the x-axis). Bottom-right: WFS noise error. Please note that the low frequency s-PSD figures and the high frequency error figure have different spatial scales: the width of the low frequency images is $5\ \text{m}^{-1}$ while the width of the high frequency s-PSD image is $20\ \text{m}^{-1}$.}
   \label{fig:06}
\end{figure}

The s-PSD profile and the corresponding PSF profile are shown in Figure \ref{fig:04}. It is well known \cite{jolissaint:01,sivaramakrishnan:03}, as can be seen with this example, that the PSF wings structure mimics the PSD shape. Figure \ref{fig:06} shows the four fundamental errors s-PSD in the spatial frequency plane. The noise and aniso-servo errors affects the central parts of the low frequency domain. Anisotropy of the aniso-servo error is a combined consequence of the wind direction and the off-axis location of the NGS. WFS aliasing, as it is expected, affects the highest spatial frequencies of the low frequency domain. 
\newpage
One can therefore expect, in general, the noise errors to contribute essentially to a widening of the PSF core, the aniso-servo error to affect the PSF wings in the region between the core and the transition to the residual seeing halo (due to the high frequency error), while aliasing would affect mostly the transition region. In this example, noise clearly dominates the PSF structure, though.

\paragraph{A note on the spatial frequency pixel size}

We have seen that both aniso-servo and aliasing s-PSD equations include cosine functions of products in $\bfr\cdot\mathbf{v}$ and $\bfr\cdot\bth$. These cosine terms need to be well sampled in the spatial frequency domain when building the numerical matrices $f_x$ and $f_y$: an under-sampling would lead to an underestimate of the wavefront error variance, as the later is estimated from numerical integration of the s-PSD, and an incorrect representation of the PSF wings structures. The consequence of such an under-sampling is an over-optimist estimate of the Strehl ratio for large off-axis NGS angle (the Strehl would saturate above a certain value while it should absolutely converge to zero for larger and larger off-axis angles). The same is true for the servo-lag error, where the performance would be over-estimated for large WFS integration time and/or high wind speed. Practically, our experience with PAOLA shows that the cosine terms should be sampled with at least 10 samples over one period. Nyquist sampling is by far not sufficient here.

\subsection{Strehl ratio of the four fundamental wavefront errors}\label{sec:42}

\begin{figure}[p]
   \centering
   \epsfig{file=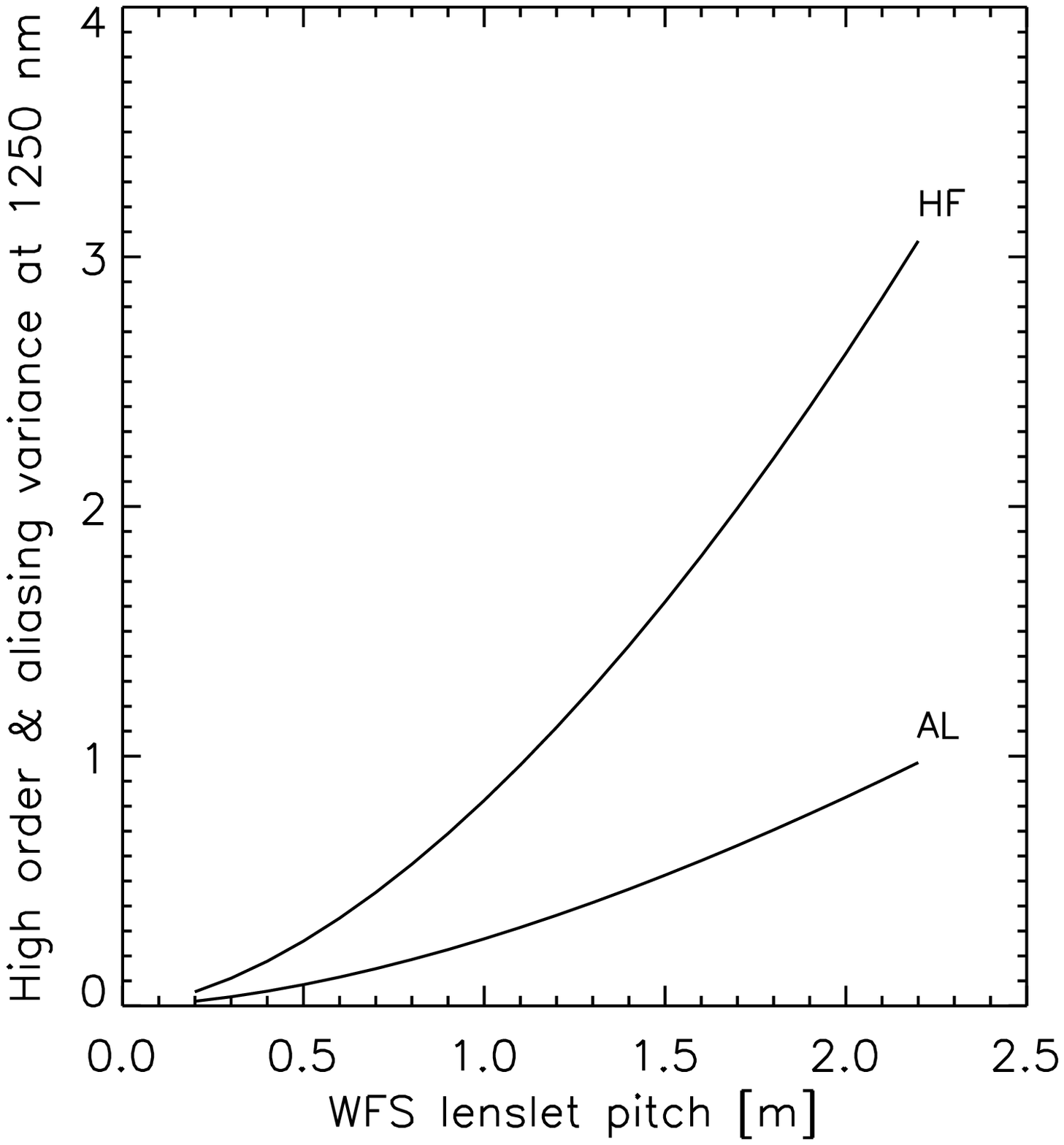,height=7.7cm,width=7.7cm}
   \hspace{2mm}
   \epsfig{file=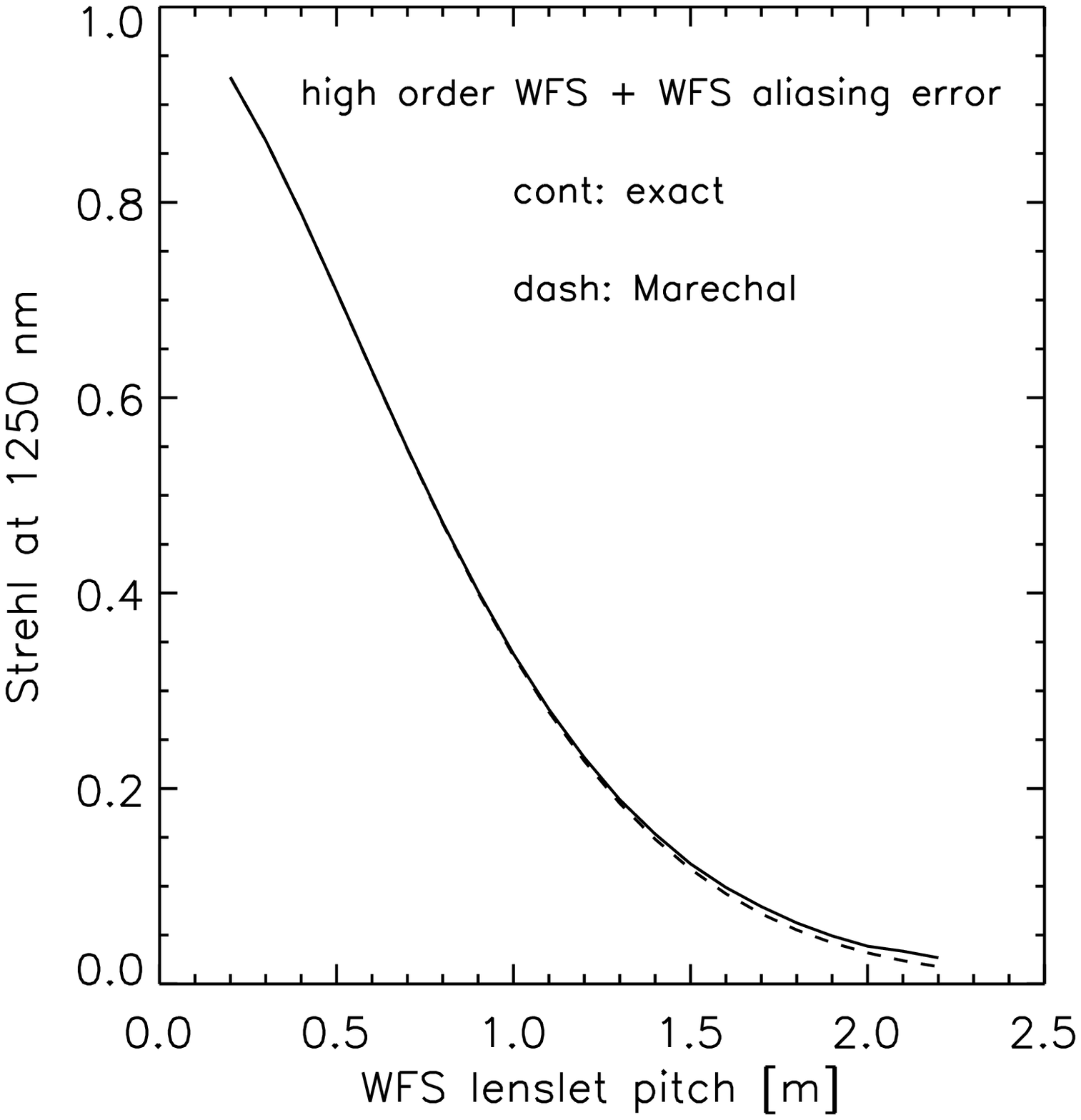,height=7.7cm,width=7.7cm}
   \caption{Left: High order WFS error and WFS aliasing phase variance as a function of the WFS lenslet pitch. It is known -- as can be seen here -- that for a SH-WFS the aliasing variance is about 1/3rd of the high frequency error. Right: Strehl ratio and WFS lenslet pitch. See section \ref{sec:42}. Dashed curves shows the Strehl computed from the Mar\'echal approximation.}
   \label{fig:07}
   \vspace{5mm}
   \epsfig{file=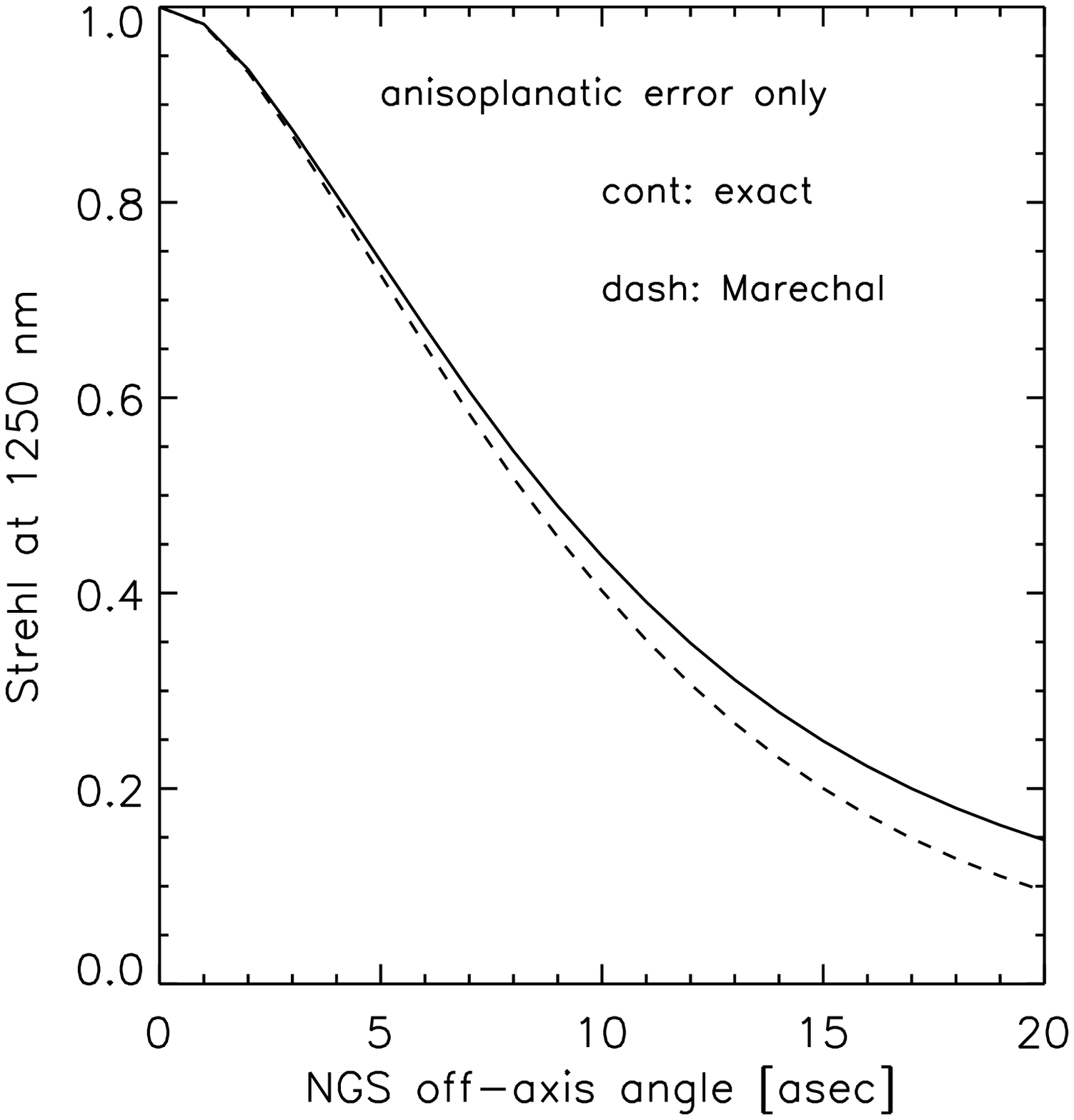,height=7.7cm,width=7.7cm}
   \hspace{2mm}
   \epsfig{file=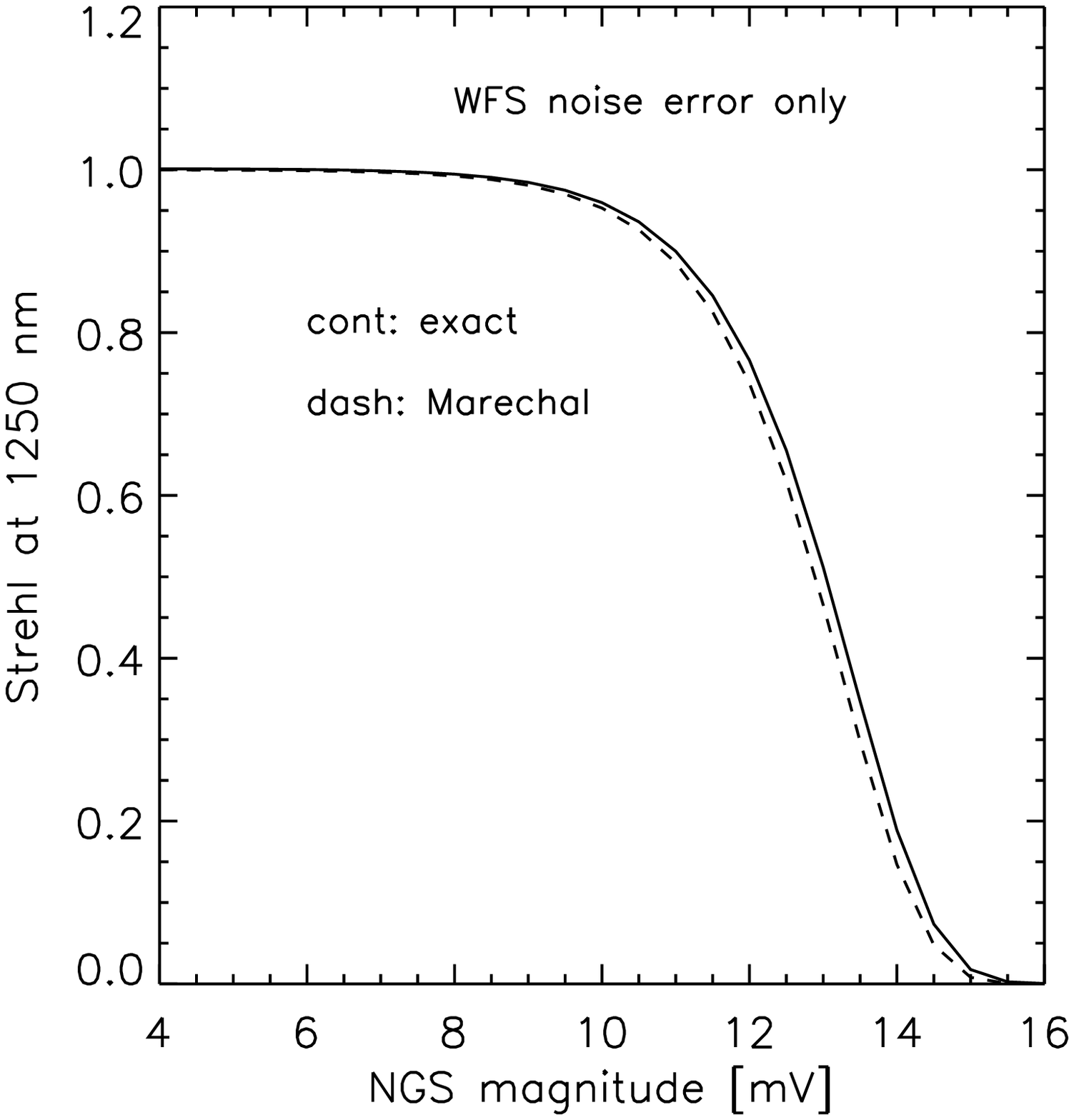,height=7.7cm,width=7.7cm}
   \caption{Left: Angular anisoplanatism Strehl and NGS off-axis angle. Dashed curves shows the Strehl computed from the Mar\'echal approximation. Right: Strehl and NGS magnitude. Dashed curves shows the Strehl computed from the Mar\'echal approximation.}
   \label{fig:09}
\end{figure}

In this section we simply illustrate how the Strehl associated with the four fundamental errors varies with the main system parameter associated to each error: the WFS lenslet pitch for the WFS high frequency and WFS aliasing error (Figure \ref{fig:07}), the NGS off-axis angle for the anisoplanatic error (Figure \ref{fig:09}, left), and the NGS magnitude for the WFS noise error (Figure \ref{fig:09}, right). It is worth noting that these curves were built in only a couple of seconds of CPU time (iMac computer, 3.06 GHz Intel Core 2 Duo processor). We tested also the Mar\'echal approximation, stating that for low to moderate phase variance $\sigma^2_\varphi$ the Strehl ratio is given by $S=\exp(-\sigma^2_\varphi)$. See the dashed curves in Figures \ref{fig:07} and \ref{fig:09}. We find that this approximation is actually excellent for the high frequency and WFS aliasing error, relatively good for the WFS noise error, and more questionable for the anisoplanatic error -- below a Strehl ratio of about 40\% in the example given here.

\subsection{Open loop versus closed loop performance}

We claimed in the introduction that open and closed loop systems behave differently in dim NGS conditions, and limiting magnitude might be quite different for both modes. This claim came from the realization that the noise transfer function are quite different in the two cases, as well as the servo-lag error transfer functions. In order to illustrate this, we computed the Strehl at 1.25 microns for our standard conditions, and a NGS magnitude in the range 4 to 16.

\begin{figure}[htb]
   \centering
   \epsfig{file=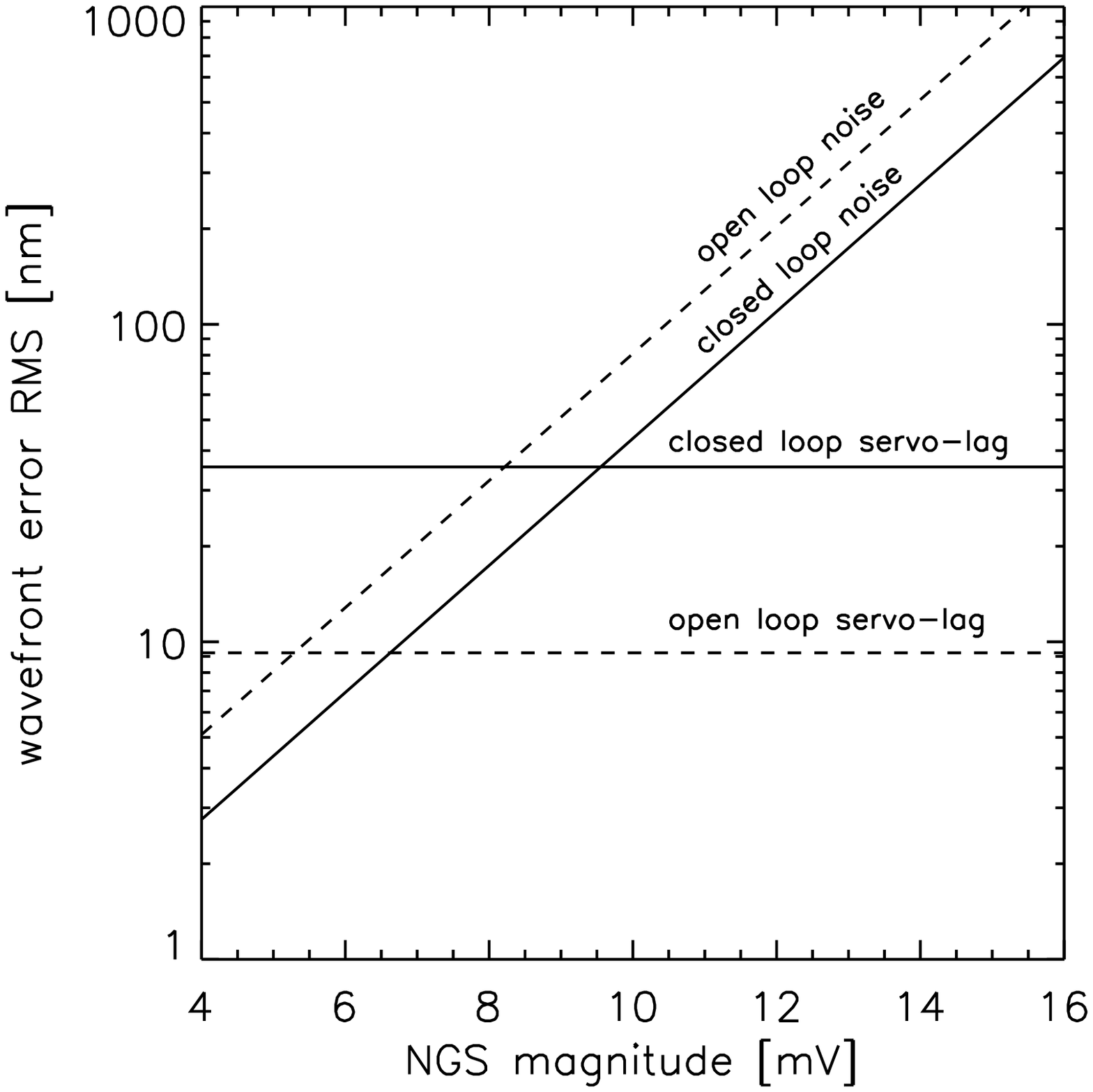,height=7.7cm,width=7.7cm}
   \hspace{2mm}
   \epsfig{file=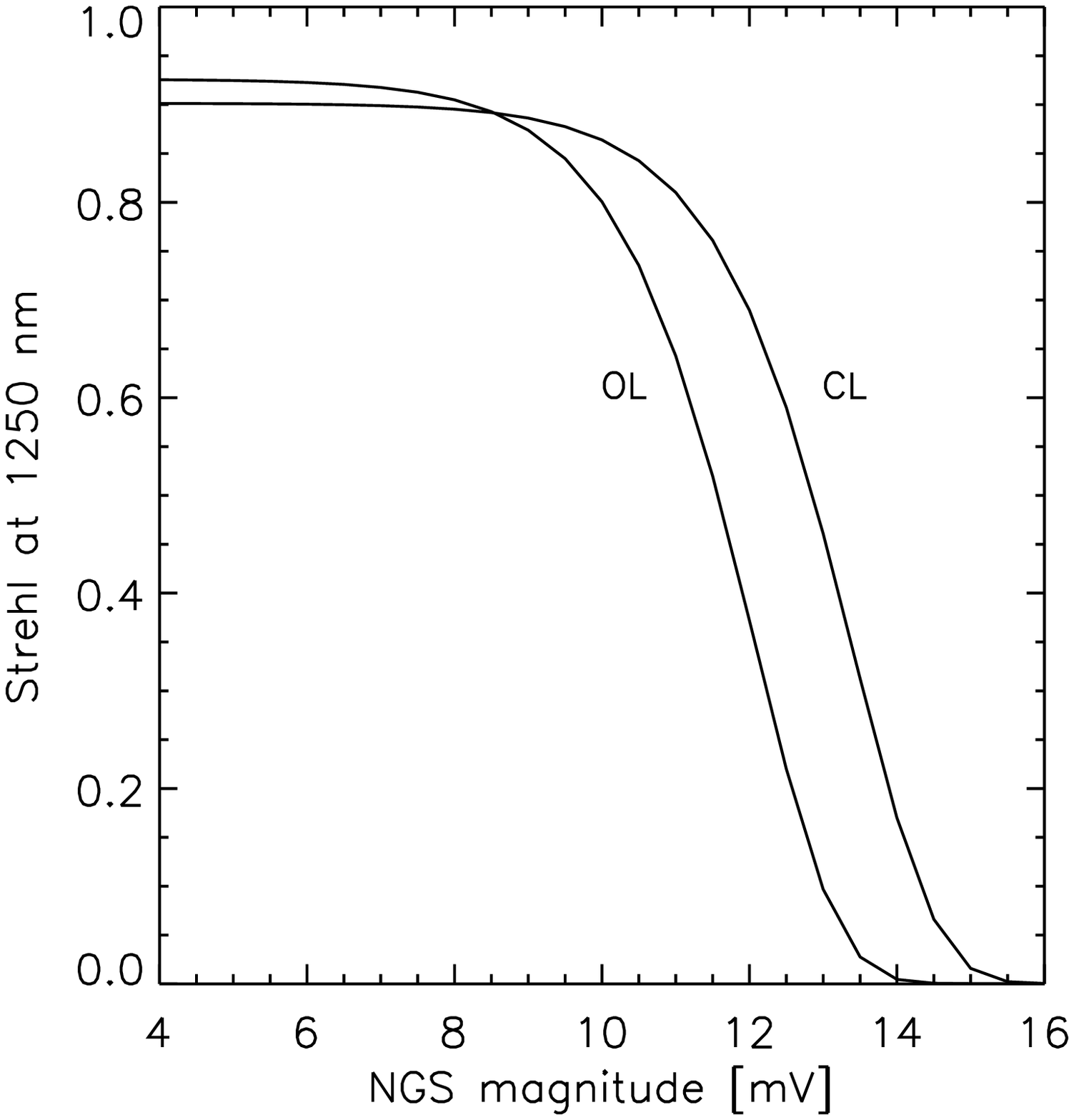,height=7.7cm,width=7.7cm}
   \caption{Left: Servo-lag and WFS noise error RMS in open and closed loop mode, for an exposure time of 1 ms and a loop gain of 0.5. Right: Strehl associated with the previous WFE, at 1.25 $\mu$m (OL: open loop, CL: closed loop).}
   \label{fig:11}
\end{figure}

\begin{figure}[htb]
   \centering
   \epsfig{file=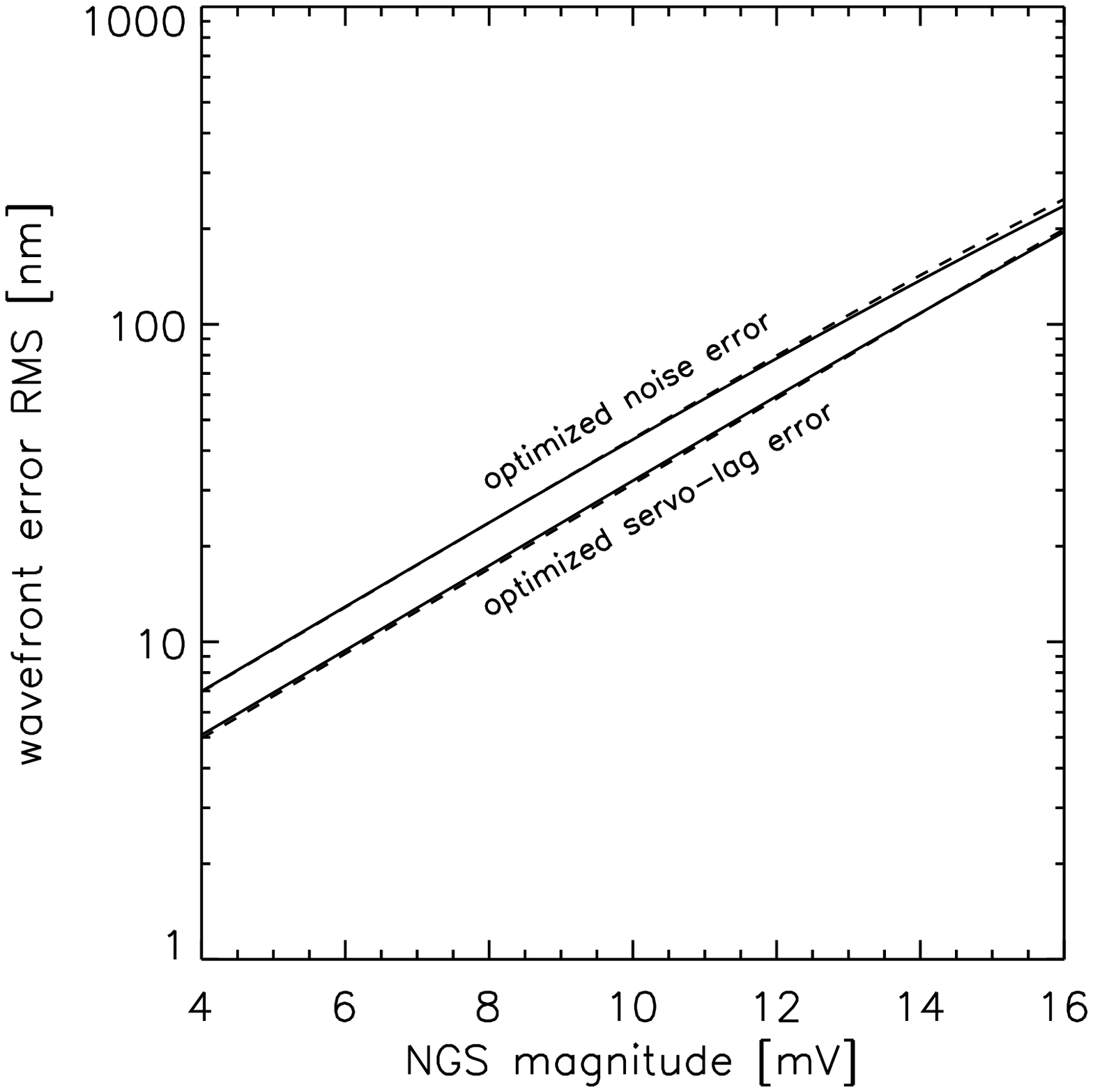,height=7.7cm,width=7.7cm}
   \hspace{2mm}
   \epsfig{file=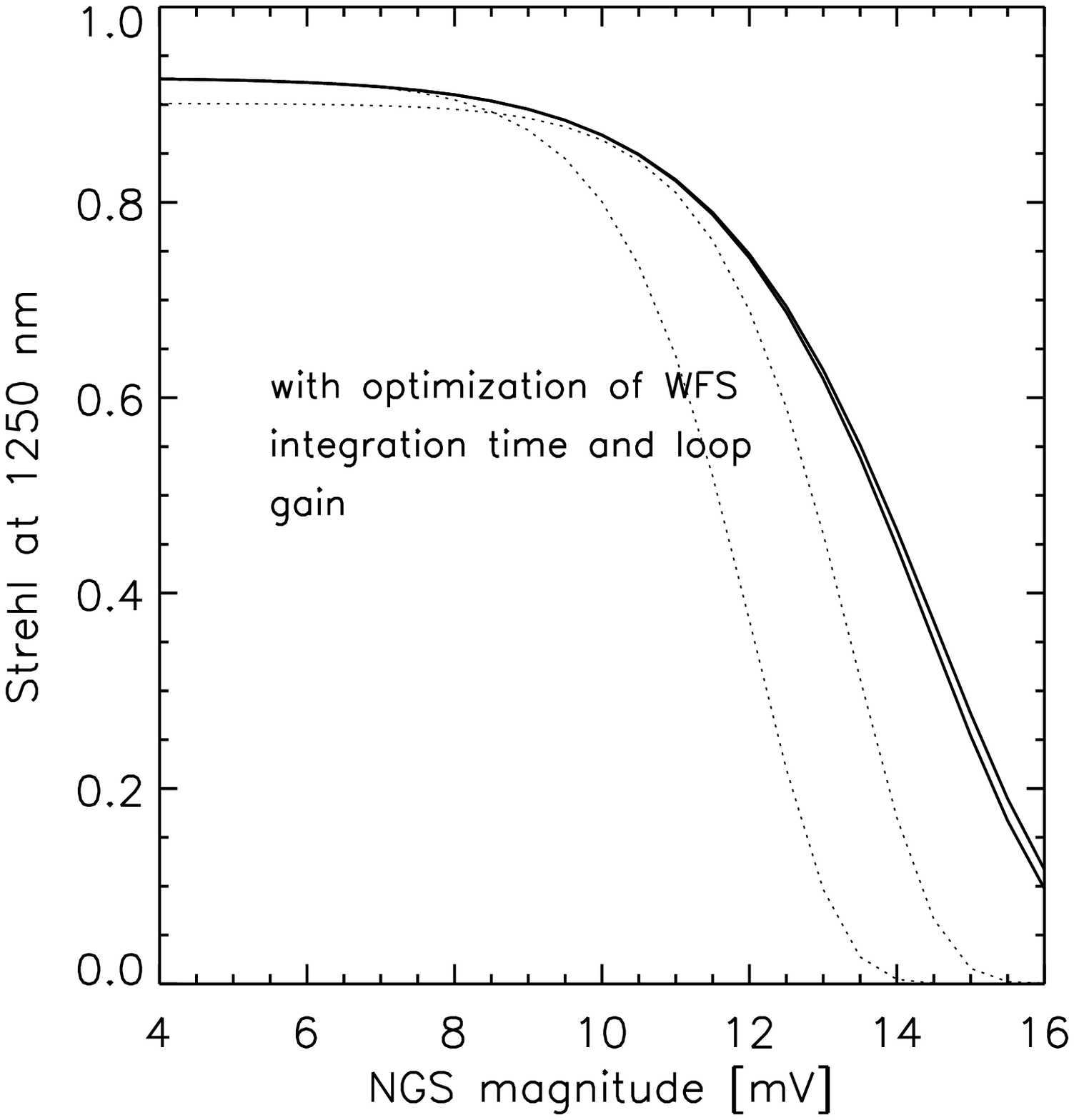,height=7.7cm,width=7.7cm}
   \caption{Left: Servo-lag and WFS noise error RMS in open and closed loop mode, same conditions than Figure \ref{fig:11}, but with optimization of the WFS integration time and loop gain. The optimized open and closed loop modes WFE are now basically indistinguishable. Right: Strehl associated with the previous WFE, at 1.25 $\mu$m; dotted line: before optimization; continuous lines: after optimization.}
   \label{fig:13}
\end{figure}

Initially, we set the WFS integration time fixed at 1 ms, for both modes, and a closed loop gain of 0.5. See Figure \ref{fig:11}. We find that the servo-lag error is higher for the closed loop mode, because the rejection transfer function has a lower bandwidth in closed loop than in open loop. The noise error on the other hand is higher in open loop, and this is because the noise is basically unfiltered in open loop, while the noise transfer function is a low pass in closed loop, filtering the high temporal frequency of the white noise spectrum. For given wind conditions, an open loop system can be run faster than a closed loop system because the servo-lag error is intrinsically lower. Therefore, we might think that a dimer NGS could be used in open loop. This is actually true only for bright NGS, where the increased open loop noise error is still low with respect to the servo-lag error. One can see for instance in Figure \ref{fig:11}, right, that for a Strehl specification of 0.9 (as it would be for an Extreme AO system), the open loop limiting magnitude would be 4 magnitudes higher than for the closed loop system. For dimer NGS, though, the increase of open loop noise overcomes the decrease of servo-loop error, and the open loop system performs less than its equivalent (same WFS integration time) closed loop system. To summarize, {\it for a given WFS integration time}, open loop systems outperform closed loop systems only in bright NGS conditions.
\newpage
As a final experiment, we optimized the WFS integration time and the closed loop gain, for each value of the NGS magnitude. See Figure \ref{fig:13}. Optimization has several consequences. First, the limiting magnitude gain is very significant, more than two magnitudes in this example. Second, it makes the open and closed loop servo and noise error converge: this is explained by the fact that the structure of the servo-lag and noise s-PSD are the same in both modes (see Figure \ref{fig:06}), therefore optimization converge towards the same solution. Note that the open loop advantage for bright stars disappears with optimization: the closed loop mode performs the same as the open loop mode at any magnitude. Therefore, contrary to what the intuition would tell, from a control efficiency point-of-view, we assert that there is no advantage of using an open loop rather than a closed loop scheme in NGS AO mode.

\section{CONCLUSION}

This paper presents a synthetic modeling method for closed loop astronomical adaptive optics, and complements earlier work on open loop modeling using the same approach. The concept of the synthetic method and its complementarity with the more classical end-to-end modeling approach is discussed extensively. The main advantages of the method is that it allows a rapid and direct modeling of the long exposure PSF without going through a long and cumbersome Monte-Carlo process Then, we give the detailed analytical calculation of the spatial power spectrum of the residual AO corrected phase, as well as the steps to go from the power spectrum to the long exposure PSF, allowing the reader to write his/her own modeling code. Dispersion of the air refractive index is included in the model, as well as the deformable mirror spatial transfer function. This method has been implemented into our AO modeling toolbox PAOLA, and used to study a few illustrative examples of the usage of the synthetic method to explore the performance of closed loop AO systems. It is found for instance that when optimizing the WFS integration time, open loop and closed loop system have basically the same performance (same limiting magnitude). Finally, it is important to recall that the foundations of the method do not depend on the type of WFS neither on the type of wavefront reconstruction method, or control algorithm. Also, the method is in principle not limited to single NGS case but can be extended, as it has been done by others, to multi-NGS and multi-DM modes. In this paper, though, we have simply considered the case of an AO system with a single NGS, for a Shack-Hartmann type WFS, a classical LSE wavefront reconstruction and a simple integrator control.

\section*{ACKNOWLEDGEMENTS}

The closed loop synthetic model was developed by the author while at the Leiden University, The Netherlands, under contract with NOVA (Nederlandse Onderzoekschool voor de Astronomie). Initial work was undertaken while at the Herzberg Institute of Astrophysics, National Research Council Canada.

\end{document}